\documentclass[aps, showpacs, floatfix, twocolumn]{revtex4}
\usepackage{graphicx}
\usepackage{epsfig}
\usepackage{epstopdf}
\usepackage{dcolumn}
\usepackage{amsmath}

\begin{document}

\title{Charm and strange quark masses and $f_{D_s}$ from overlap fermions}

\author{
\small Yi-Bo Yang$^{1,2}$, Ying Chen$^{1}$, Andrei Alexandru$^{3}$, Shao-Jing Dong$^{2}$, Terrence Draper$^{2}$, \mbox{Ming Gong$^{1,2}$}, Frank X. Lee$^{3}$, Anyi Li$^{4}$, Keh-Fei Liu$^{2}$, Zhaofeng Liu$^{1}$, and Michael Lujan$^{5}$
\vspace*{-0.5cm}
\begin{center}
\large{
\vspace*{0.4cm}
\includegraphics[scale=0.20]{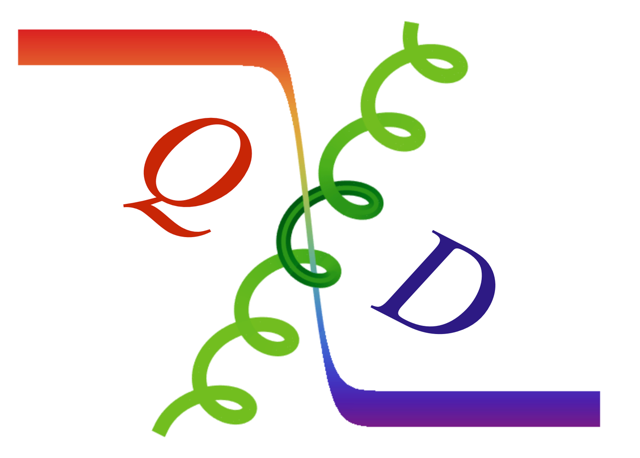}\\
\vspace*{0.4cm}
($\chi$QCD Collaboration)
}
\end{center}
}
\affiliation{
\vspace*{0.5cm}
$^{1}$Institute of High Energy Physics and Theoretical Physics Center for Science Facilities,\\
Chinese Academy of Sciences, Beijing 100049, China\\
$^{2}$Department of Physics and Astronomy, University of Kentucky, Lexington, KY 40506, USA\\
$^{3}$Department of Physics, The George Washington University, Washington, DC 20052, USA\\
$^{4}$Institute for Nuclear Theory, University of Washington, Seattle, WA 98195, USA \\
$^{5}$Math Modeling Group, LMI, Tysons Corner, VA 22102, USA \\
}

\begin{abstract}{
We use overlap fermions as valence quarks to calculate meson masses in a wide quark mass range on
the $2+1$-flavor domain-wall fermion gauge configurations generated by the RBC and UKQCD
Collaborations. The well-defined quark masses in the overlap fermion formalism and the clear valence
quark mass dependence of meson masses observed from the calculation facilitate a direct derivation
of physical current quark masses through a global fit to the lattice data, which incorporates
$O(a^2)$ and $O(m_c^4a^4)$ corrections, chiral extrapolation, and quark mass interpolation. Using the
physical masses of $D_s$, $D_s^*$ and $J/\psi$ as inputs, Sommer's scale parameter $r_0$ and
the masses of charm quark and strange quark in the $\overline{\rm MS}$ scheme are determined to be
$r_0=0.465(4)(9)$ fm, $m_c^{\overline{\rm MS}}(2\,{\rm GeV})=1.118(6)(24)$ GeV (or $m_c^{\overline{\rm MS}}(m_c)=1.304(5)(20)$ GeV), and $m_s^{\overline{\rm
MS}}(2\,{\rm GeV})=0.101(3)(6)\,{\rm GeV}$, respectively. Furthermore, we observe that the mass
difference of the vector meson and the pseudoscalar meson with the same valence quark content is
proportional to the reciprocal of the square root of the valence quark masses. The hyperfine splitting of
charmonium, $M_{J/\psi}-M_{\eta_c}$, is determined to be 119(2)(7) MeV, which is in good
agreement with the experimental value. We also predict the decay constant of $D_s$ to be
$f_{D_s}=254(2)(4)$ MeV. The masses of charmonium $P$-wave states $\chi_{c0}, \chi_{c1}$ and $h_c$ are also in good agreement
with experiments.
 }
\end{abstract}

 \maketitle

\section{Introduction}

A large endeavor has been devoted by lattice QCD to determine the quark masses which are of great
importance for precision tests of the Standard Model of particle physics~\cite{Bazavov:2009bb,Bazavov:2014wgs,Allton:2008pn,Aoki:2009ix,
Davies:2009ih,Allison:2008xk,Durr:2010vn,McNeile:2010ji,Laiho:2011np,Blum:2010ym,Kelly:2012uy,Arthur:2012opa,Carrasco:2014cwa,Heitger:2013oaa}. In the lattice QCD formulation, quark masses are dimensionless bare parameters and their renormalized values at
a certain scale should be determined through physical inputs. For the light $u,d$ quarks and the strange
quark, their masses are usually set by the physical pion and kaon masses as well as the decay
constants $f_\pi$ and $f_K$, where the chiral extrapolation is carried out through chiral
perturbation theory~\cite{Bazavov:2009bb,Allton:2008pn,Aoki:2009ix}. 
For heavy quarks, the bare quark masses are first set in the vicinity of the
physical region and the physical point can be interpolated or extrapolated through the quark mass
dependence observed empirically from the simulation. In the above procedures, non-perturbative
quark mass renormalization is usually required to match the bare quark mass to the renormalized
one at a fixed scale. For the heavy quark, the HPQCD collaboration~\cite{Allison:2008xk} proposed a promising scheme to obtain their
masses from current-current correlators of heavy quarkonium, which is free of the quark mass
renormalization~\cite{McNeile:2010ji}.

In this work we propose a global-fit strategy to determine the strange and charm quark masses which
incorporates simultaneously the $O(a^2)$ correction, the chiral extrapolation, and the strange/charm
quark interpolation. The lattice setup is a mixed action formalism where we use the overlap
fermions as valence quarks and carry out the calculation on the domain-wall gauge configurations generated by
the RBC and UKQCD Collaborations. Both the domain-wall fermion (DWF) and the overlap fermion are chiral fermions; as such, they do not have $O(a)$ errors for the valence quark masses, and the additive renormalization for them is also negligible ($10^{-9}$) due to the overlap fermion implementation. It is also shown that the nonperturbative renormalization via chiral Ward identities or the regularization independent/momentum subtraction (RI/MOM) scheme can be implemented relatively
easily. We have explored this strategy and found that it is feasible for valence
masses reaching even the charm quark region on the set of DWF configurations that we work on.

The RBC and UKQCD Collaborations have simulated $2+1$ flavor full QCD with
dynamical domain-wall fermion (DWF) on several lattices in the last decade with pion masses as low as
$\sim 300$ MeV and volumes large enough for mesons ($m_\pi L
>4$)~\cite{Allton:2008pn,Blum:2010ym,Arthur:2012opa}. It turns out that the fermions in this formalism with a finite fifth dimension $L_s$ satisfy the
Ginsparg-Wilson relation reasonably well and the chiral symmetry breaking effects can be
absorbed in the small residual masses. As for the overlap fermion, its
multi-mass algorithm permits calculation of multiple quark propagators covering the range from very
light quarks to the charm quark. This makes it possible to study the properties of charmonium and
charmed mesons using the same fermion formulation for the charm and light quarks. Having multiple masses helps
in determining the functional forms for the quark mass dependence of the observables. In practice,
we calculate the masses of charmonia and charm-strange mesons with the charm and strange quark mass
varying in a range, through which a clear observation of the valence quark mass dependence of
meson masses can be obtained. Similar calculations are carried out on six configuration
ensembles and the results are treated as a total data set for the global fit as mentioned above. It
should be noted that the quark masses in the global fit are matched to the renormalized quark
masses at 2 GeV in the minimal-subtraction scheme ($\overline{\rm MS}$ scheme) by the quark mass
renormalization constant $Z_m$ calculated in Ref.~\cite{Liu:2013yxz}. In order to convert the quantities
on the lattice to the values in physical units, we take the following prescription. First, the
ratio of the Sommer scale parameter $r_0$ to the lattice spacing $a$, namely $r_0/a$, is
measured precisely from each gauge ensemble. Subsequently, $r_0/a$'s in the chiral limit are used
to replace the explicit finite-$a$ dependence. Instead of determining the exact value of $r_0$ by a
specific physical quantity, we treat it as an unknown parameter and determine it along with the
quark masses through the global fit with physical inputs.

One of our major observations is that the masses of the pseudoscalar and the vector mesons have
clear contributions from the term proportional to the reciprocal of the square-root of the valence
quark masses, as predicted by a study based on a potential model of the quarkonium where this kind of
contribution is attributed to the scaling behavior of the spin-spin contact interaction of the valence
quarks~\cite{Liu:1977et}. This is also in quantitative agreement with the feature of the meson
spectrum from experiments. After incorporating this kind of mass dependence to the global fit, the
experimental value of the hyperfine splitting of the $1S$ charmonium, the mass difference of
$J/\psi$ and $\eta_c$, can be well reproduced after the charm quark mass, the strange quark mass,
and $r_0$ at the physical point are determined by using $J/\psi$, $D_s^*$ and $D_s$ masses as input. 
We also extract the decay constant $f_{D_s}$
of the $D_s$ meson both from the partially conserved axial current relation and the direct
definition of $f_{D_s}$ along with the renormalization constant $Z_A$ of the axial vector current.
The two derivations give consistent results which are also in agreement with the experimental value
within errors. The masses of charmonium $P$-wave states $\chi_{c0}, \chi_{c1}$ and $h_c$ are also predicted and they
are in good agreement with experiments.

This work is organized as follows. We give a detailed description of our numerical study in
Section~\ref{sec:numerical}, where we focus on the derivation of $r_0/a$ and its chiral
extrapolation, the quark mass renormalization, and the investigation of the valence quark mass dependence of mesons, particularly the hyperfine splitting.
The global fit details and the major results on quark masses and $f_{D_s}$ are given in Section~\ref{sec:results}, where a thorough discussion of the statistical and systematic errors is also presented. The summary and the conclusions are presented in Section~\ref{sec:summary}.

\section{Numerical details}\label{sec:numerical}
 Our calculation is carried out on the $2+1$ flavor domain wall fermion configurations generated by the RBC/UKQCD
 Collaborations~\cite{Aoki:2010dy}. We use two lattice setups, namely, the $L^3\times T = 24^3\times 64$ lattice
 at $\beta=2.13$ and the $32^3\times 64$ lattice with $\beta=2.25$. For the $\beta=2.13$ lattice,
 the mass parameter of the strange sea quark is set to $m_s^{(s)}a=0.04$ and that of the
 degenerate  $u/d$ sea quark takes the values of $m_l^{(s)}a=0.005, 0.01$, and 0.02, which give three
 different gauge ensembles. 
  However, it is found that the physical strange quark mass parameter is
  actually $m_s^{(s)}a=0.0348$~\cite{Aoki:2010dy} as determined by the physical $\Omega$ baryon mass, this 
 discrepancy has been corrected by the corresponding reweighting factors. 
 Similarly, the $m_s^{(s)}a$ of the $\beta=2.25$ lattice
 is set to 0.03 in generating the three gauge ensembles with $m_l^{(s)}a$ taking values $0.004,
 0.006$, and $0.008$, but the physical $m_s^{(s)}a$ 
 for $\beta=2.25$  is found to be 0.0273~\cite{Aoki:2010dy}. 
 Since the physical values of the sea quark masses are not the same on the two sets of configuration with
 different $\beta$, we shall assess its systematic error by introducing a linear $m_s^{(s)}$ dependent term in 
 the global fitting formula and observing the effects when the sea strange mass is shifted to the physical values determined by
 global fitting itself.
 On the other hand, the explicit
 chiral symmetry breaking of the domain wall fermions gives rise to the residual mass $m_{\rm res}a$
 for the sea quarks which has been studied by RBC and UKQCD~\cite{Aoki:2010dy}. These corrections to the light sea quark masses are taken into account for the chiral limit. The parameters of the six gauge ensembles involved in this work  are listed in Table~\ref{table:para}, and the numbers of configurations we used are listed in Table~\ref{table:Ncfg}.
 
\begin{table}[htbp]
\begin{center}
\caption{\label{table:para}The parameters for the RBC/UKQCD configurations~\cite{Aoki:2010dy}.
$m_s^{(s)}a$ and $m_l^{(s)}a$ are the mass parameters of the strange sea quark and the light sea quark,
respectively. $m_{\rm res}^{(s)}a$ is the residual mass of the domain wall sea quarks.}
\begin{ruledtabular}
\begin{tabular}{ccccccc}
$\beta$ & $L^3\times T$  & $m_s^{(s)}a$ &   &  $m_l^{(s)}a$  &    &  $m_{\rm res}^{(s)}a $ \\
\hline
2.13    &$24^3\times 64$ & 0.04   &0.005&0.01    &0.02  &  0.00315(4)      \\
2.25    &$32^3\times 64$ & 0.03   &0.004&0.006   &0.008 & 0.00067(1)
\end{tabular}
\end{ruledtabular}
\end{center}
\end{table}

\begin{table}[htbp]
\begin{center}
\caption{\label{table:Ncfg}The number of configurations of the six ensembles used in this work.}
\begin{ruledtabular}
\begin{tabular}{c|ccc|ccc}
&  & $24^3\times 64$ & &&$32^3\times 64$ &\\
$m_l^{(s)}a$ & 0.005&0.01    &0.02&0.004&0.006   &0.008\\
$n_{cfg}$ & 99& 107&100 &53&55&50\\
\end{tabular}
\end{ruledtabular}
\end{center}
\end{table}

We use the overlap fermion action for the valence quarks to perform a mixed-action study in this
work. The massless overlap fermion operator $D_{\rm ov}$ is defined as 
\begin{equation}
D_{\rm ov}=1+\gamma_5\epsilon(H_W(\rho)),
\end{equation}
where $\epsilon(H_W(\rho))$ is the sign function of the Hermitian matrix $H_W(\rho)=\gamma_5
D_W(\rho)$ with $D_W(\rho)$ the usual Wilson-Dirac operator with a negative mass parameter $-\rho$.
Thus the effective massive fermion propagator $D_c^{-1}(ma)$ can be defined through $D_{\rm ov}$ as~\cite{Chiu:1998eu, Liu:2002qu}
\begin{equation}\label{massive}
D_c^{-1}(ma)\equiv \frac{1}{D_c(0) + ma}, \textrm{ with } D_c =\frac{\rho D_{\rm ov}}{1-D_{\rm ov}/2}
\end{equation}
where $ma$ is the bare mass of the fermion. From the Ginsparg-Wilson relation $\{\gamma_5, D_{\rm
ov}\}=\rho aD_{\rm ov}\gamma_5 D_{\rm ov}$, one can check the relation $\{\gamma_5, D_c(0)\}=0$~\cite{Chiu:1998gp}, which
implies that the mass term $ma$ in Eq.~(\ref{massive}) acts the same way as an additive term to
the chirally-invariant Dirac operator in
the continuum Dirac operator and thus there is no additive mass renormalization. On the other hand, in
order for the chiral fermion to exist, $\rho$ should take a value in the range $0<\rho<2$, so we
take the optimal value $\rho=1.5$ which gives the smallest $(ma)^2$ error in the hyperfine splitting and the fastest production of $D_c^{-1}(ma)$~\cite{Li:2010pw}. Through the multi-mass algorithm, quark propagators $D_c^{-1}(ma)$ for two dozen
different valence quark masses $ma$ have been calculated in the same inversion, such that we can calculate
the physical quantities at each valence quark mass and obtain clear observation of the quark mass
dependence of these quantities.

\subsection{The ratio of the Sommer scale and the lattice spacing}
The unique dimensionful parameter in the lattice formulation of QCD is the lattice spacing $a$,
which is usually determined through a sophisticated scheme. 
Although dimensionful quantities, such as $f_{\pi}, f_K$, and hadron masses have been used
to determine the lattice spacing, the lattice results need to be extrapolated to the continuum limit and physical
pion mass in order for the experimental values to be used as inputs for such a determination. 
In contrast to 
the hadronic quantities which have explicit dependence of quark masses, the Sommer parameter,
$r_0$ (or $r_1$), which is relatively easy to calculate, has been used to set the scale. 
Still, it needs to be determined precisely at the chiral and continuum limits. $r_0$ is defined by the relation~\cite{Sommer:1993ce},
\begin{equation}\label{r0:def}
\left[r^2\frac{dV(r)}{dr}\right]_{r=r_0}=1.65,
\end{equation}
where $V(r)$ is the static potential in the heavy quark limit ($r_1$ is defined similarly with 1.65
replaced by 1~\cite{Aubin:2004wf}). Practically in each gauge ensemble, $V(r)$ can be derived from
the precise measurement of Wilson loops $W(r,t)$ with different spatial and temporal extensions
$(r,t)$ as
\begin{equation}
\langle W(r,t)\rangle \sim e^{-V(r)t}.
\end{equation}

\begin{figure}[htbp]
\includegraphics[scale=0.40]{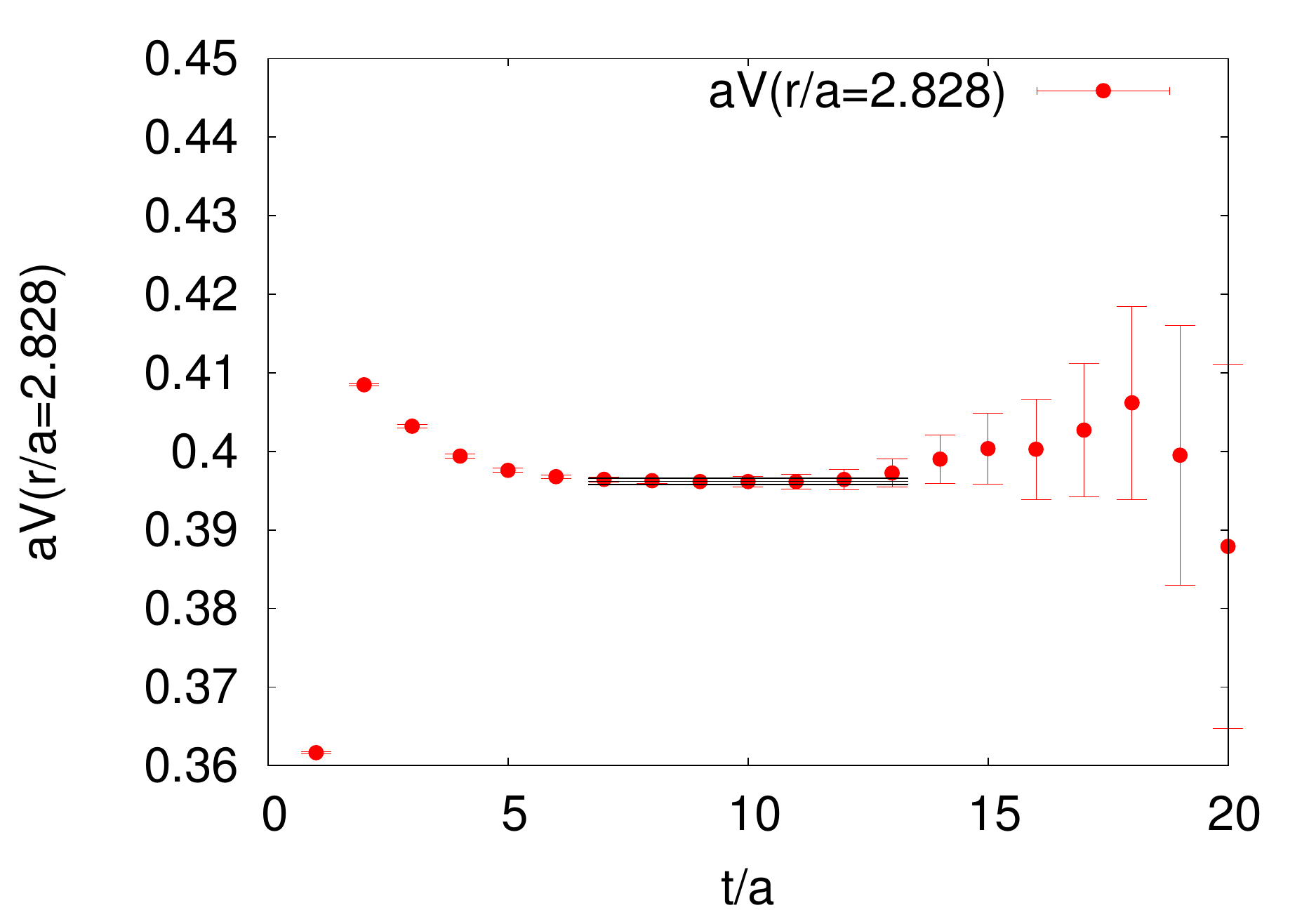}
\includegraphics[scale=0.40]{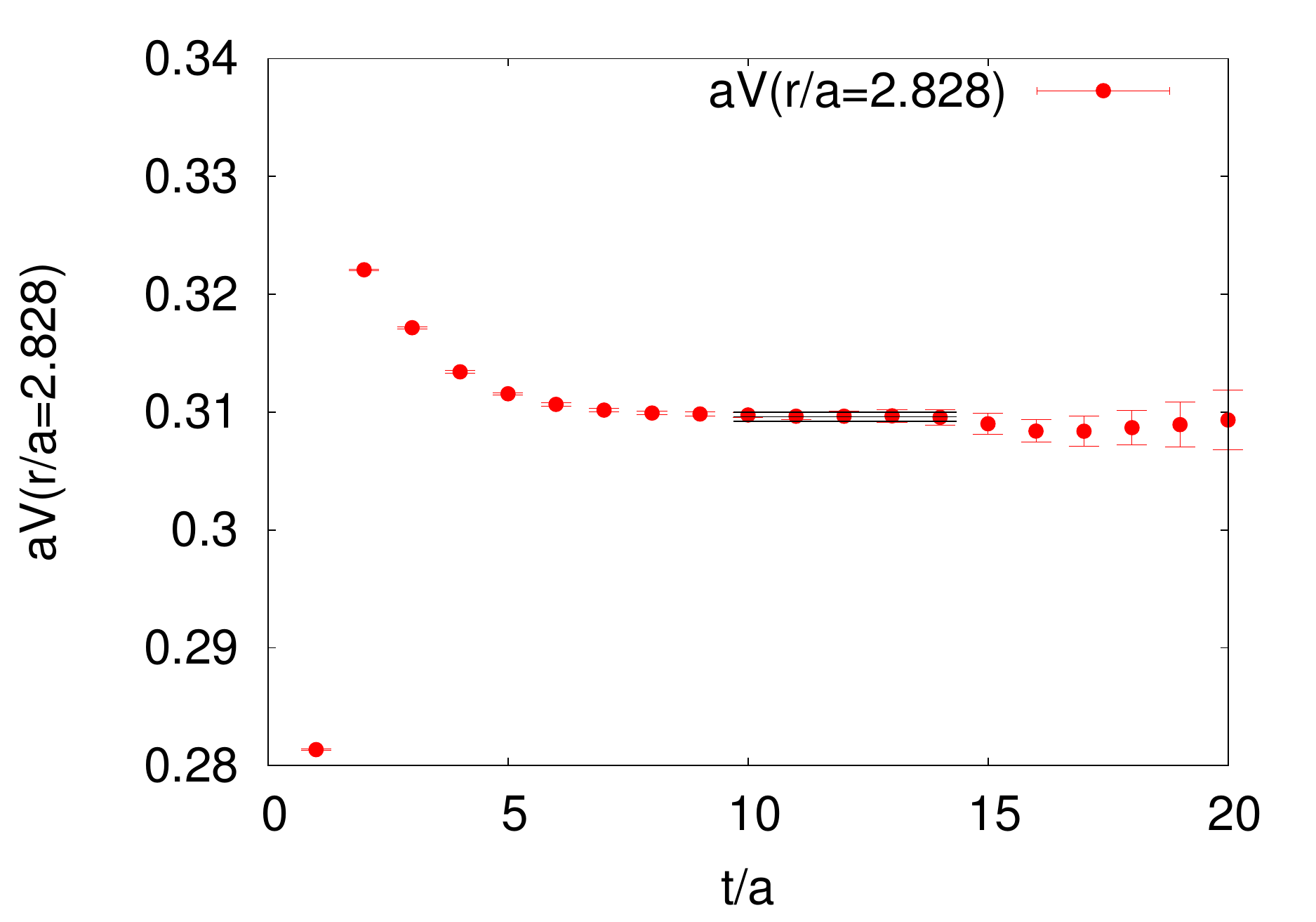}
\caption{\label{fig:v(r)}The plateau of heavy quark potential in coarse/fine lattice ensembles with
lightest sea quark masses. The upper panel is for the $24^3\times 64$ lattice with $m_l^{(s)} a=0.005$, the lower panel for the
$32^3\times 64$ lattice with $m_l^{(s)} a=0.004$}
\end{figure}

Fig.~\ref{fig:v(r)} shows the effective plateaus of $V(r)$ at $r/a=2.828$ with respect to $t/a$
. One can see the measurements are very precise
and the plateaus last long enough (from 8 to 15 approximatively) for a precise determination of $r_0/a$.

$V(r)$ is usually parametrized in the Cornell potential form,
\begin{equation}
V(r)=V_0-\frac{e_c}{r}+\sigma r,
\end{equation}
where $\sigma$ is the string tension. Considering the lattice spacing $a$ explicitly, the potential
one measures on the lattice is actually
\begin{equation}    \label{potential}
a V(r/a) = aV_0 -\frac{e_c}{r/a} + (\sigma a^2)  r/a,
\end{equation}
and the $aV_0$, $e_c$, and $\sigma a^2$ can be obtained from a correlated minimal-$\chi^2$ curve
fitting to $\langle V(r,t)\rangle$'s through Eq.~(\ref{potential}). 
For each gauge ensemble, one can find the ratio
\begin{equation}
\frac{r_0}{a}=\sqrt{\frac{1.65-e_c}{\sigma a^2}}
\end{equation}
using Eq.~(\ref{r0:def}). Table~\ref{table:r0} lists the calculated $r_0/a$'s for the six ensembles we are
using. Note that the sea quark masses (both light and strange) are bare quark masses of the domain wall
fermion, the physical ones should include the residual masses (0.00315(4) and 0.00067(1)
for the two lattices respectively~\cite{Aoki:2010dy}) and the mass renormalization factor (1.578(2) and 1.527(6) correspondingly~\cite{Aoki:2010dy}) in the $\overline{MS}$
scheme at 2 GeV, i.e.
\begin{equation}  \label{eq:renormalized_mass}
m^{(s),R}_{l}a = Z_m^{(s)} (m^{(s)}_{l}+m_{res}^{(s)}) a
\end{equation}

\begin{table}[htbp]
\begin{center}
\caption{\label{table:r0} $r_0/a$'s and the sea quark masses renormalized in $\overline{MS}$ scheme at 
2 GeV for the six ensembles (EN1, EN2 and EN3 at each of $\beta$) we are using. 
The residual masses of the light domain wall sea quark have been included in the sea quark masses. The extrapolated values at the physical point (3.408(48) MeV) ~\cite{Bazavov:2009bb,Durr:2010vn,McNeile:2010ji,Laiho:2011np,Blum:2010ym,Kelly:2012uy} are also listed. See the text for more details.}
\begin{ruledtabular} 
\begin{tabular}{cc|ccc|c}
&  & EN1 & EN2 & EN3 & physical point\\
\hline
$\beta=2.13$ &$m^R_la$ & 0.03653      &  0.02075       & 0.01286     &   --     \\
             &$r_0/a$& 3.906(3)  & 3.994(3)    & 4.052(3)  & 4.114(10)\\
\hline
$\beta=2.25$ &$m^R_la$ & 0.01323     & 0.01018       & 0.00713     & --       \\
             &$r_0/a$& 5.421(5)  & 5.438(6)    & 5.459(6)  & 5.494(3)

\end{tabular}
\end{ruledtabular}
\end{center}
\end{table}

\begin{figure}[htbp]
\includegraphics[scale=0.7]{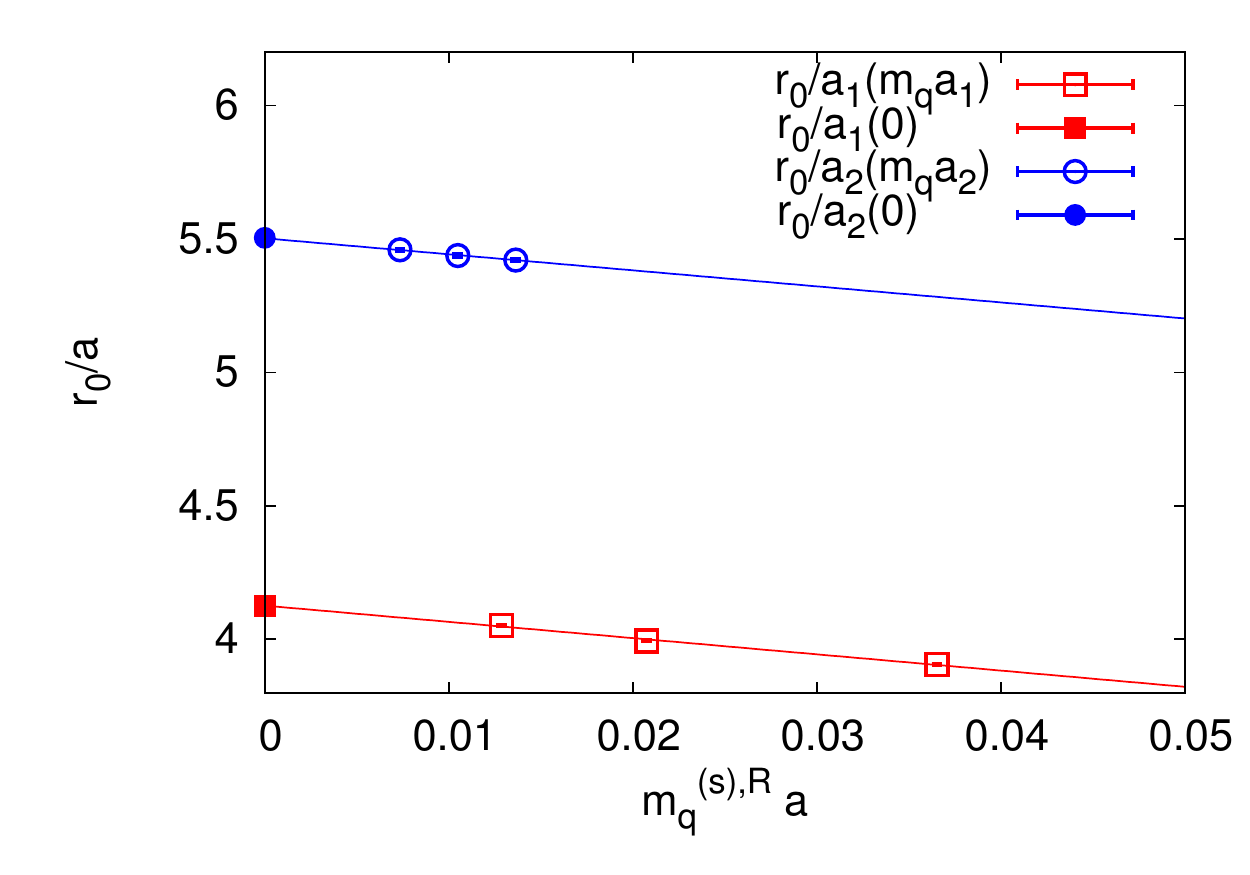}
\caption{Renormalized sea quark mass $m^R_q a$ dependence of $r_0/a$. Square points for $\beta=2.13$ with lattice spacing $a_1$, 
and circular points for $\beta=2.25$ with lattice spacing $a_2$.}\label{fig:r0_2}
\end{figure}

 The $r_0$ dependence on the lattice spacing $a$ and the sea quark mass up to $O(a^2)$
can be expressed as~\cite{Davies:2009tsa, Arthur:2012opa}
\begin{eqnarray}\label{eq:r0_ori}
r_0(a,m_l,m_s)&=&r^0_0(1+\sum_i c^{a}_i a^{2i})\nonumber\\
&+&(c^{(l)}+d^{(l)} a^2)(m^{R}_{l}-m^{phys}_{l})\nonumber\\
&+&(c^{(s)}+d^{(s)} a^2) (m^{R}_{s}-m^{phys}_{s}).
\end{eqnarray}
Note that the sea quark masses $m_l$ and $m_s$ should take the renormalized mass values at an energy
scale in Eq.~(\ref{eq:r0_ori}) in order for the $c,d$ coefficients 
in the equation to be free of the $a$-dependence.
For the ensembles with the same $\beta$, the behavior of $r_0/a$ with respect to the light sea
quark mass $m_l^{(s)} a$ is shown in Fig.~\ref{fig:r0_2}, where the square points are for the coarse lattice $\beta=2.13$
($24^3\times 64$ lattices), and the circular points are for the fine lattice $\beta=2.25$ ($32^3\times 64$ lattices). 

In this work, we don't determine $r_0$ (or  the lattice spacing) before the fit of the value of interest like the hadron masses and decay constants. Instead, we will use three hadronic quantities to obtain the $r_0$ (and also $m_s$ and $m_c$) with 
\begin{equation}\label{eq:r0_over_a}
 C(a)\equiv \frac{r_0}{a}(m_l^{phys},a)
\end{equation} 
as the extrapolated value of a linear fit in 
$m^R_la$ for each lattice, i.e.
\begin{equation}\label{eq:r0_fit}
\frac{r_0}{a}(m^R_la,a)=\frac{r_0}{a}(m^{phys}_l,a)+f^{(l)}(a)(m^{R}_l-m^{phys}_l)a.
\end{equation}
Without the information of the lattice spacing, we have to do the fit with extrapolating $r_0/a$ to the chiral limit to produce a value of $r_0$ and also the lattice spacings of the ensembles at $\beta=2.13/2.25$, then we can extrapolate $r_0/a$ to $m_l^{phys}=3.408(48)$MeV coming from the lattice average and iterate the fit until the $r_0/a$ is converged. The extrapolated values $C(a)$
are also listed in Tab.~\ref{table:r0}.

Through such a fit, we get $f^{(l)}(a)=6.08(44)$ for
$\beta=2.13$ and $f^{(l)}(a)=6.19(36)$ for $\beta=2.25$, which are independent of the
lattice spacing within errors. This implies that the coefficient $c^{(l)}$ in Eq.~\ref{eq:r0_ori} is very small and
 is consistent with zero. The strange sea quark mass has been tuned to be close to the physical point,
so we ignore the strange sea quark mass dependence and treat the effect due
to deviation from the physical point as a source of the systematic uncertainty.

In view of the above discussion for the subsequent global fits, we shall use the following fitting form
\begin{equation}\label{eq:r0_new}
r_0(a,m_l,m_s)=r^0_0(1+
c^a_1 a^2)+d^l a^2(m^R_l-m_l^{phys}).
\end{equation}

\subsection{Quark mass renormalization}\label{subsec:massrenorm}

 In lattice QCD, the bare quark masses are input parameters in lattice units, say, $m_q a$. However in the global
fit including the continuum extrapolation to be discussed later, $m_qa$ has to be converted 
 to the renormalized current quark mass $m_q^{R}(\mu)$ at a fixed scale $\mu$ and a fixed scheme (usually $\overline{MS}$) 
which appears uniformly in the global fitting formulas for different lattice spacings. This requires the renormalization 
constant $Z_m$ of the quark mass for each $\beta$ to
be settled beforehand.

When we use the chirally regulated field $\hat{\psi}=(1-\frac{1}{2}D_{\rm ov})\psi$ in the definition of the 
interpolation fields and currents for the overlap fermion, the renormalization constants of scalar ($Z_S$), 
pseudoscalar ($Z_{P}$), vector ($Z_V$), and axial vector ($Z_A$) currents obey the relations $Z_S=Z_P$ and
$Z_V=Z_A$ due to chiral symmetry. In addition,
$Z_m$ can be derived from $Z_S$ by the relation $Z_m=Z_S^{-1}$. In Ref.~\cite{Liu:2013yxz}, the RI-MOM scheme is adopted 
to do the non-perturbative renormalization on the lattice to obtain the renormalization constants under
that scheme; those values are then converted from the RI-MOM scheme to the $\overline{\rm MS}$ scheme using
ratios from continuum perturbation theory. The relations between $Z$'s mentioned above are verified, and the renormalization 
constants obtained are listed in Table~\ref{table:renorm_const}.
Besides the statistical error, systematic errors including those from the scheme matching and the running of quark masses in the
 $\overline{\rm MS}$ scheme are also considered in
Ref.~\cite{Liu:2013yxz}. The systematic error from the running quark mass in the $\overline{\rm
MS}$ scheme is negligibly small, while the one from scheme matching is at four loops, and has a size of about
1.4\%. The errors of $Z_S^{\overline{\rm MS}}(2\,{\rm GeV})$ quoted above include both the statistical and systematic ones.
A systematic discussion on the
renormalization of overlap fermion on domain wall fermion sea is given in Ref.~\cite{Liu:2013yxz}. 

\begin{table}[htbp]
\begin{center}
\caption{\label{table:renorm_const} The renormalization constants of the overlap fermion on domain wall fermion sea. Both the statistical error and the systematic errors from scheme matching and
running of quark masses in the $\overline{\rm MS}$ scheme are considered.}
\begin{ruledtabular}
\begin{tabular}{c|cc}
& $Z_S$ & $Z_m$ \\
\hline
$\beta=2.13$ &  1.127(9)(19)     & 0.887(7)(15)     \\
\hline
$\beta=2.25$ & 1.056(6)(24)       & 0.947(6)(20)   \\
\end{tabular}
\end{ruledtabular}
\end{center}
\end{table}

With the above prescriptions, we can replace the renormalized quark masses and $a$ by the bare
quark mass parameters $m_q a$,  $r_0$, $Z_m(2\,{\rm GeV},a)$ and $C(a)$ defined in Eq.~(\ref{eq:r0_over_a}) as
\begin{equation}\label{mass_convert}
m_q^R(2\,{\rm GeV}) = Z_m(2\,{\rm GeV},a) (m_qa) \frac{C(a)}{r_0}.
\end{equation}
In this way $r_0$ enters into the global fitting formula to be discussed in Sec.~\ref{sec:quark_mass_dep}
as a new parameter and can be fitted
simultaneously with other parameters in the fitting formulas.

The discretization errors of the renormalization constant $Z_m$ could induce an extra $ma^2$ term
on the quark mass dependence of a given meson mass. Note that the lattice spacings used in Ref.~\cite{Liu:2013yxz} are slightly different from those which will be obtained in this paper, and it could be considered as a source of the $ma^2$ dependence.

\subsection{The quark mass list}\label{sec:quark_mass_usage}

Since we are using the overlap fermion action for valence quarks, we can take advantage of the
multi-mass algorithm with little computation overhead to calculate the valence quark
propagators for dozens of different quark masses $ma$ in the same inversion. Subsequently, multiple
quantities can be calculated at these valence quark masses, such that their quark mass dependence
can be clearly observed. Because we have not determined the concrete values of lattice spacings yet, we
first estimate the meson masses in the strange and the charm quark mass regions using the approximate values $a^{-1}\sim 1.75 $ GeV for $\beta=2.13$ and $a^{-1}\sim 2.30$ GeV for $\beta=2.25$ as determined by RBC and UKQCD~\cite{Aoki:2010dy} where both the sea and valence quarks 
are domain-wall fermions. 
We obtain the dimensionless masses of the pseudoscalar and vector mesons for different valence quark masses 
through the relevant two-point functions which are calculated with the $Z_3$ grid source to increase statistics~\cite{Li:2010pw}.
The physical strange quark mass is estimated to be around $m_sa=0.06$ for $\beta =2.13$ and
$m_sa=0.04$ for $\beta=2.25$; thus we choose the $m_sa$ region to be $m_sa\in[0.0576,0.077]$ and
$m_sa\in[0.039,0.047]$ for the two lattices, respectively. We cover a wider range for the charm quark mass, i.e.
$ [0.29,0.75]$ and $[0.38,0.57]$ for the two lattices to study the charmonium and charm-strange mesons. The concrete
strange and charm quark mass parameters in this work are listed in
Table~\ref{table:quark_mass}.
\begin{table}[htbp]
\begin{center}
\caption{The bare mass parameters for valence strange and charm quarks in this
study.}\label{table:quark_mass}
\begin{ruledtabular}
\begin{tabular}{c|c|lllllll}
$\beta=2.13$ & $m_sa$ & 0.0576 & 0.063 & 0.067 & 0.071 & 0.077 &      &     \\
             & $m_ca$ & 0.29   & 0.33  & 0.35  & 0.38  & 0.40  & 0.42 & 0.45\\
             &        & 0.48   & 0.50  & 0.53  & 0.55  & 0.58  & 0.60 & 0.61    \\
             &        &   0.63  & 0.65  & 0.67  & 0.68  & 0.70  & 0.73  & 0.75    \\
             &        &    &    &   &    &    &   &     \\
$\beta=2.25$ & $m_sa$ & 0.039  & 0.041 & 0.043 & 0.047 &       &      &     \\
             & $m_ca$ & 0.38   & 0.46  & 0.48  & 0.50  & 0.57  &      &

\end{tabular}
\end{ruledtabular}

\end{center}
\end{table}

\begin{figure}[htbp]
\includegraphics[scale=0.7]{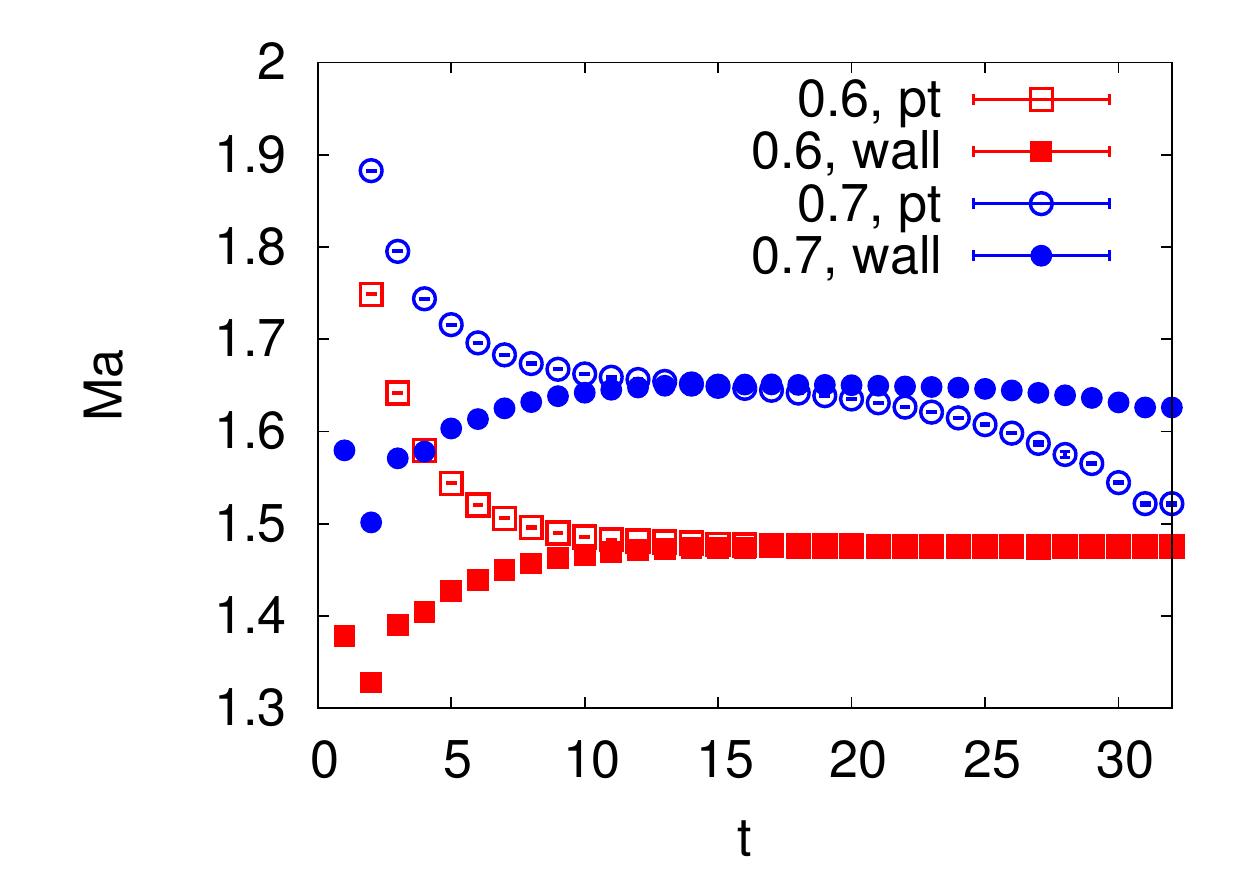}
\caption{Effective mass plot of the pseudoscalar meson mass ($Ma$) at fixed quark masses ($ma$).
The plateau in the $ma$=0.7 case with point source is not reliable. The one with coulomb gauge fixed wall source is better, while the plateau still drops down for the $t>25$ region.  This plot is based on the result from
 the gauge ensemble at $\beta=2.13$ with the light sea quark mass
$m_l^{(sea)}a=0.05$.
}\label{fig:eff_mass_ps}
\end{figure}

It should be noted that for $\beta = 2.13$ at physical charm quark mass, discretization artifacts prevent us from computing charmonium states' masses using point-source propagators. For $a m_c \ge 0.7$, the quark propagators receive an unphysical contribution related to the locality radius of the overlap operator. This can be seen from an heavy quark expansion of the propagator:
$$
\frac1{D+m} \approx \frac1m \left[ 1 - \frac Dm + \left(\frac Dm\right)^2+\ldots\right] \,. 
$$
The off-diagonal (non-local) elements of the operator $D$  decay exponentially like $e^{-r/r_{(0)}}$ \cite{Draper:2005mh}. For large masses the quark propagator will be dominated by the first term in the expansion, $D/m$, and at large distances the decay rate will be set by $1/r_{(0)}$ rather than $m$. This regime should set in around the point where the quark mass is comparable with $1/r_{(0)}$. For $\beta = 2.13$ the locality radius for the quark bilinear state is about $1.5a$. In Fig. 3 we plot the effective mass for the pseudoscalar $\bar{c}c$ state using both point sources and wall-sources. For $am=0.7$ both propagators show signs of this unphysical state at large times, but for the wall-source propagator the effect of the unphysical state is weaker and the effective mass forms a plateau whereas for the point source it never plateaus. We use the wall-source propagators to extract the masses for charmonium states.

We have used two-term fitting to account for the effect of the excited state. But for safety, we have to use the coulomb wall source propagator to construct the charmonium correlators in the three ensembles with $\beta=2.13$ since the physical $m_c a$ is around 0.7. Note that we continue to use point source propagator for the $\bar{c}s$ system to obtain the decay constant of $D_s$. It should be safe since the unphysical mode is much heavier than $M(D_s)$ or $M(D_s^*)$. In the case of the $\beta=2.25$ ensembles, the correlators based on the point source propagators are used and the standard 
interpolation of the physical charm quark is applied, since the physical $m_c a$ is around 0.5 and does not suffer from the problem of the unphysical state.

\subsection{The quark mass dependence of meson masses}\label{sec:quark_mass_dep}

\begin{figure}[htpb]
\begin{center}
  \includegraphics[scale=0.70]{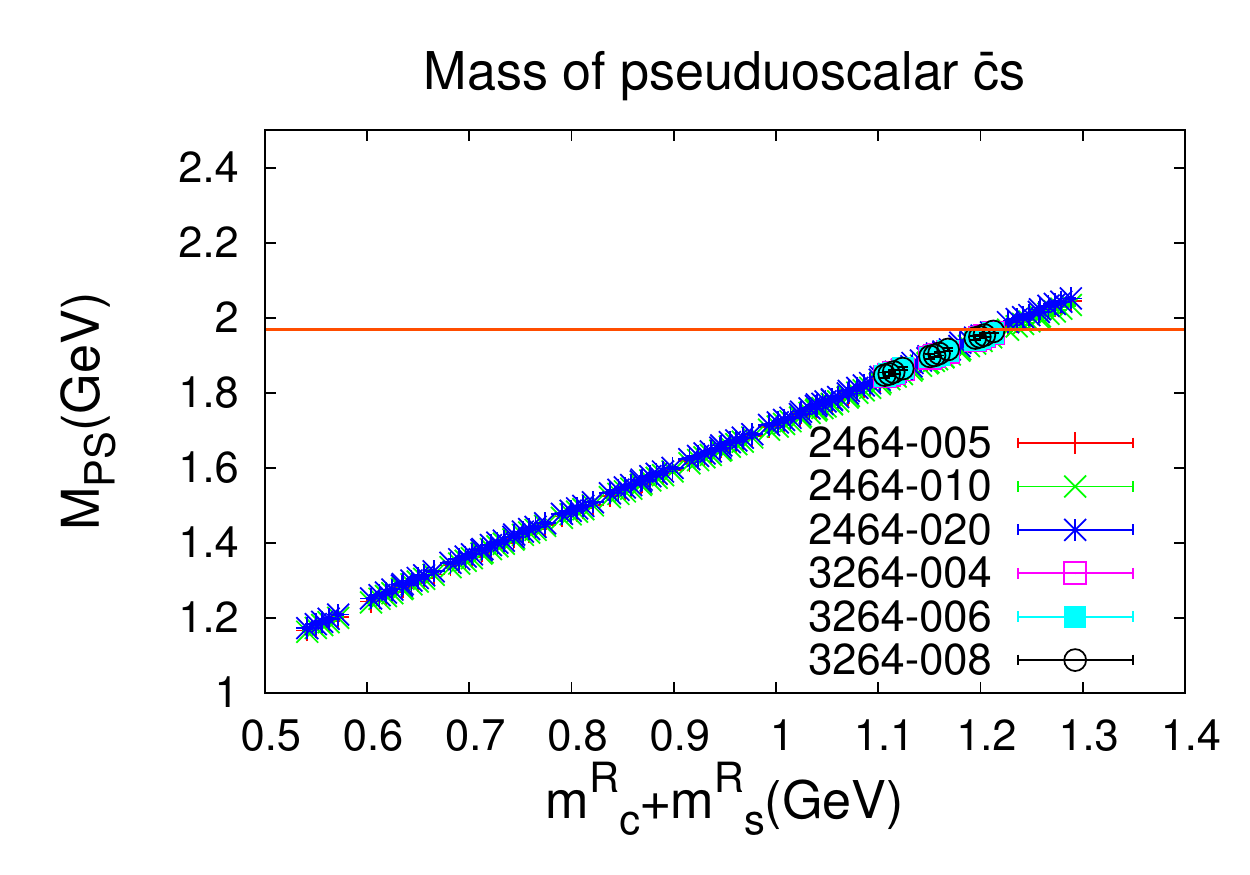}
  \includegraphics[scale=0.70]{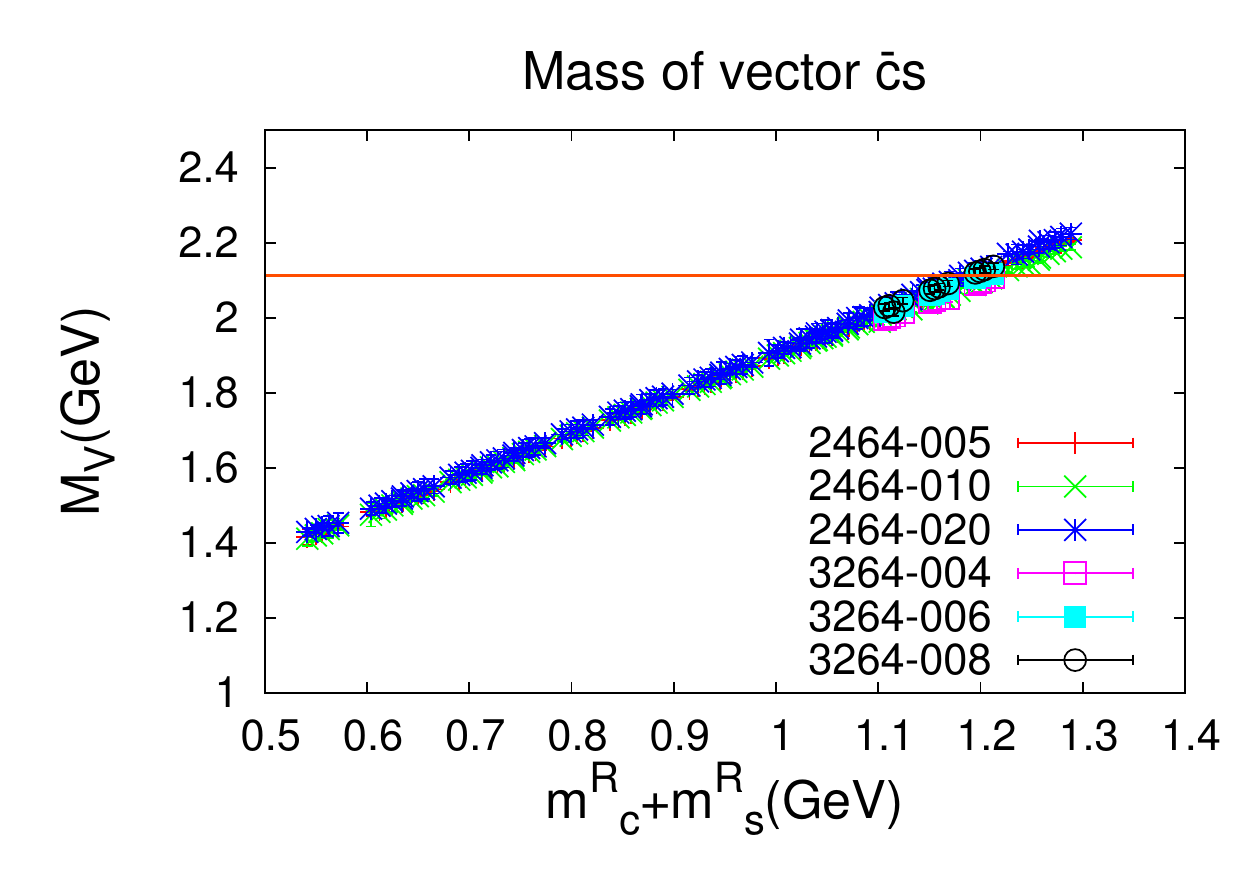}
\caption{\label{fig:cs_mesons}The masses of the pseudoscalar and vector $\bar{c}s$ mesons are plotted with respect to the renormalized 
$m_c^R+m_s^R$ (tentatively taking $r_0=0.46$)
for the six RBC/UKQCD gauge ensembles, where the linear behaviors
in $m_c^R+m_s^R$are clearly seen. The horizontal lines in the plot are the physical value of 
$D_s$ in the upper panel and $D_s^*$ in the lower panel.}\label{fig:cs_S}
\end{center}
\end{figure}

\begin{figure}[htpb]
\begin{center}
  \includegraphics[scale=0.70]{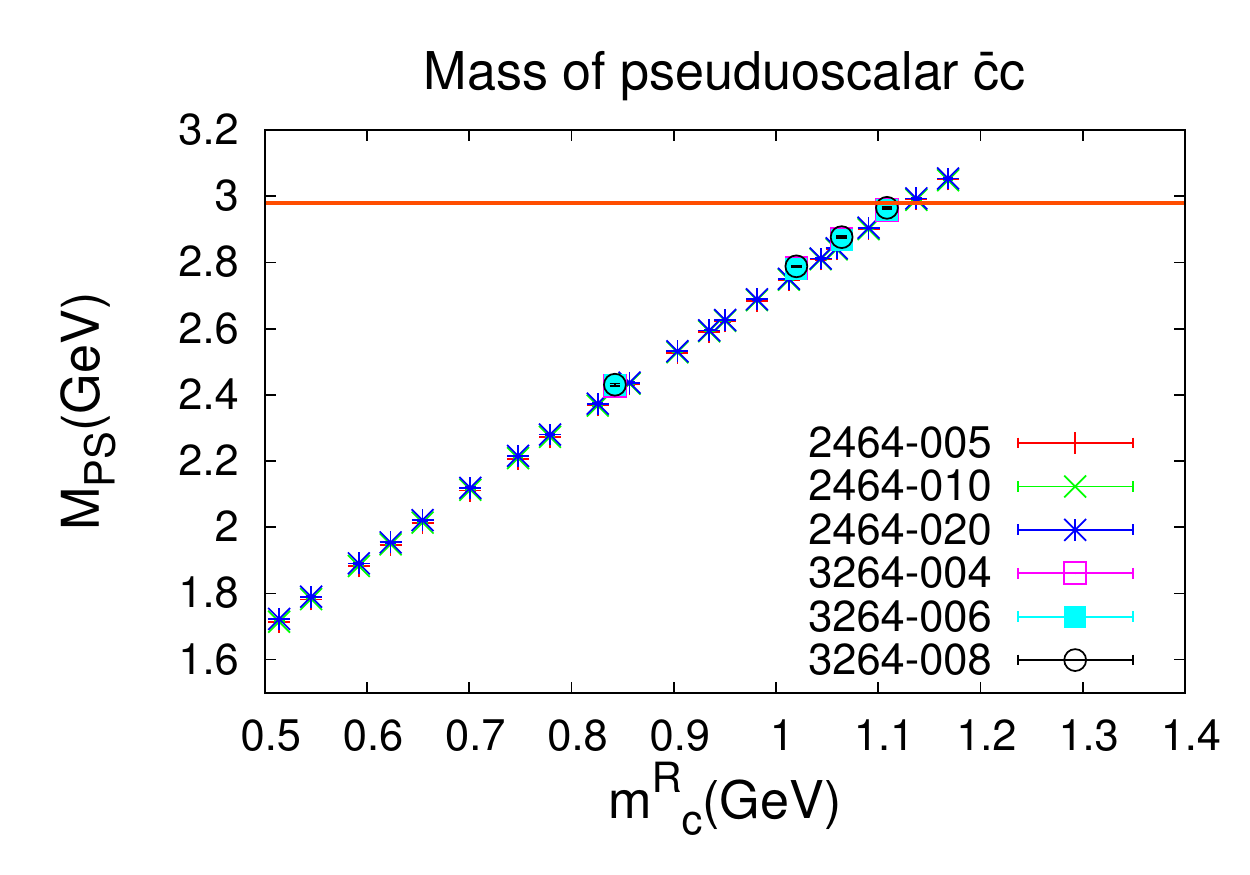}
  \includegraphics[scale=0.70]{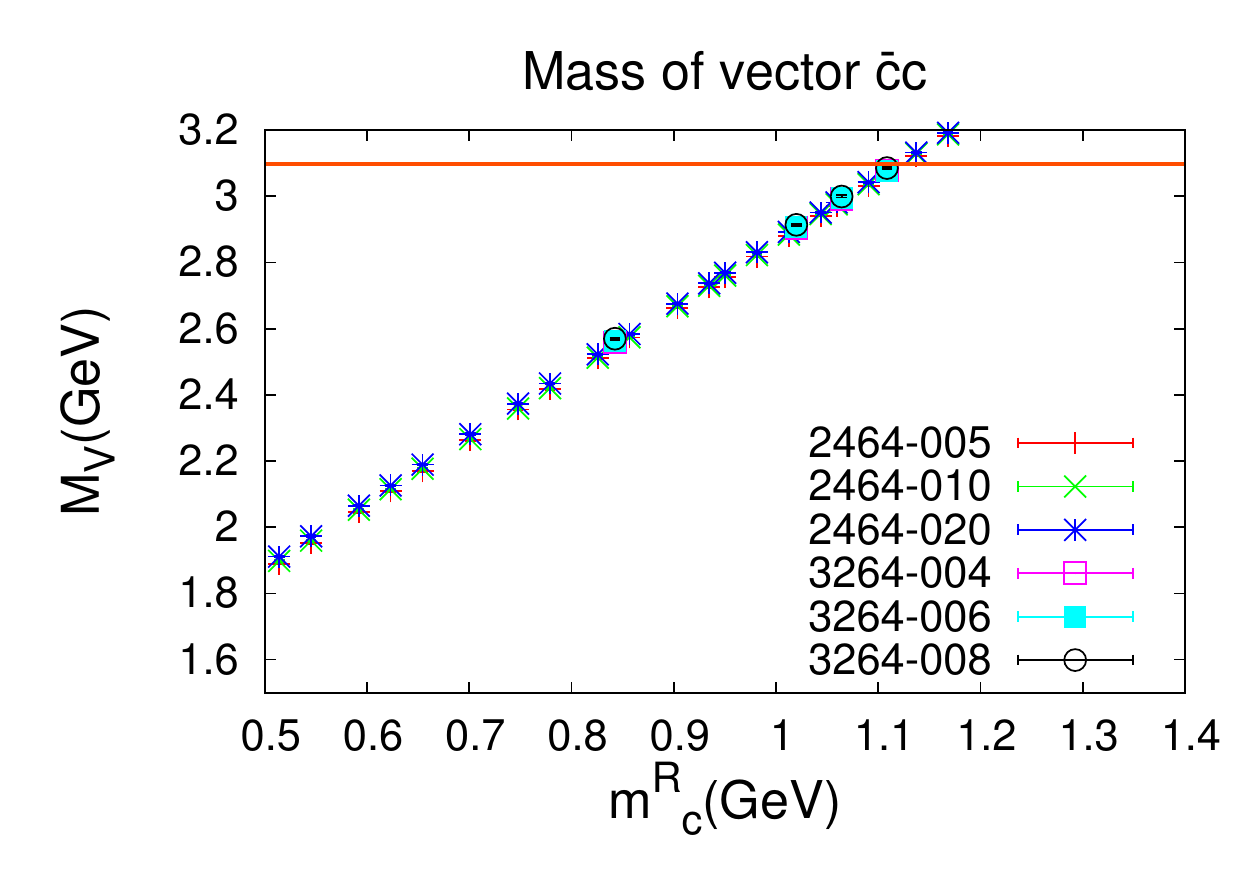}
\caption{\label{fig:cc_mesons}The quark mass dependence of the masses of the pseudoscalar and vector $c\bar{c}$ mesons is illustrated in the plots for the six RBC/UKQCD gauge ensembles, where the
linear behaviors in $m_c^R$ (tentatively taking $r_0=0.46$) are clearly seen. 
The horizon lines in the plot are the physical value $\eta_c$ in the upper panel and
$J/\Psi$ in the lower panel.}\label{fig:cc_S}
\end{center}
\end{figure}

For each of the six gauge ensembles, we calculate the masses of the pseudoscalar and the vector mesons of
the $c\bar{s}$ and $c\bar{c}$, with the strange and charm quark taking all the possible
values in Table~\ref{table:quark_mass}. Fig.~\ref{fig:cs_mesons} shows the quark mass dependence
of $c\bar{s}$ mesons, where the upper panel is for the pseudoscalar ($D_s$) and the lower panel is
for the vector ($D_s^*$). The abscissa is the sum of the renormalized strange and charm quark
masses at $2$ GeV in the $\overline{MS}$ scheme, which are converted through
Eq.~(\ref{mass_convert}) by tentatively taking $r_0=0.46$ fm, for example. Meson masses are also converted into
values in the physical unit using this scale parameter. Note that we are focusing on the behavior at the
moment, instead of the precise values of the masses here. It is interesting that the $D_s$ and $D_s^*$ masses
are almost completely linear in $m_c^R+m_s^R$ for both lattices. The light sea quark mass dependence is
very weak for $D_s$ masses in the upper panel of Fig.~\ref{fig:cs_mesons} but sizable for $D_s^*$ masses
in the lower panel. On the other hand, the slopes with
respect to $m_c^R+m_s^R$ are approximately the same for $D_s$ and $D_s^*$, 
while they still slightly depend on the lattice spacing. 
The red horizontal lines in the figure show the physical $D_s$ and $D_s^*$
masses, and the intersection regions with the data indicate where the physical $m_c$ and $m_s$
should be. Fig.~\ref{fig:cc_mesons} is similar to Fig.~\ref{fig:cs_mesons}, but for $\eta_c$ and
$J/\psi$, where one can see the similar feature of the charm quark mass dependence of $\eta_c$ and
$J/\psi$ masses. 

Based on the above observations, we assume tentatively dominance by linear dependence of meson masses
on the quark masses,
\begin{equation}\label{eq:para-form2}
M^{(0)}(m_c,m_s,m_l)= A_0 +A_1 m_c +A_2 m_s  +A_3 m_l + \ldots,
\end{equation}
where the coefficients $A_i$ can be different for different mesons, but are independent of the
lattice spacing and quark masses, since $m_c, m_s, m_l$ here are the current quark masses in the continuum QCD Lagrangian, which can be defined at an energy scale through a renormalization scheme and are independent of the lattice
spacing $a$. 

Although Fig. 4 and 5 suggest that the $a$-dependence is mild, it is incorporated with the usual generic formula
 of the charm quark mass dependence of the physical observables for the
charmonium and charm-light mesons on the lattice. It is expressed as~\cite{Davies:2010ip}
\begin{eqnarray}\label{eq:para-form}
&&M(m_c, m_s, m_l, a)=M(m_c,m_s,m_l)\nonumber\\
&\times&(1+B_1 (m_ca)^2+B_2(m_ca)^4+O((m_ca)^6 ))\nonumber\\
&+& C_1 a^2+O(a^4).
\end{eqnarray}
With the help of chiral perturbation theory, $M(m_c,m_s,m_l)$ in Eq.~(\ref{eq:para-form}) could be 
the theoretical function for the quantity $M$ which is better known for light quarks. But
for charmonium and charm-strange mesons, the functional form is not well-known
but can be investigated empirically from the lattice observations such as that of Eq.~(\ref{eq:para-form2}). The 
polynomial with respect to $m_ca$ in the parentheses takes into account the lattice artifacts of the
lattice quark actions. Since we use chiral fermion actions both for the sea quarks (domain wall
fermions) and the valence quarks (overlap fermions), chiral symmetry guarantees that they are
automatically improved to $O(a^2)$ and higher order artifacts due to the heavy quark
 are even powers of $m_ca$~\cite{Davies:2010ip}. In the ensembles with $\beta=2.13$, even the effect of the $m_c^4a^4$ term could be important since the $m_c a$ of the physical charm quark is around 0.7. It motivates us to use a large number of quark mass parameter values in those ensembles to determine this effect precisely.
There should be also similar terms for $m_l$ and $m_s$, but they are much smaller in
comparison with that of $m_c$ and can be neglected. Also included in
Eq.~(\ref{eq:para-form}) is the explicit artifact in terms of $a^2$ which comes from the lattice gauge
action and other sources of $a$-dependence.

However, this does not complete the investigation of the valence quark mass dependence of the meson
masses. For example, if we apply the functional form above in the correlated fit of the mass of $D_s$, the $\chi^2/d.o.f.$ is 4.6, much larger than unity. 

The reason is simple. Given the linear behavior described above, it would be expected that the mass difference of
the vector and the pseudoscalar mesons with the same flavor content is also proportional to the
valence quark mass. But the experimental results give a different story: for example,
$M_\rho-M_\pi\sim 630$ MeV, $M_{K^*}-M_K\sim 400$ MeV, $M_{D^*}-M_{D}\sim 140$ MeV,
$M_{D_s^*}-M_{D_s}\sim 140$ MeV, $M_{J/\psi}-M_{\eta_c}\sim 117$ MeV, etc. do not have
a linear dependence in the sum of their constituent quark masses. 

This motivates us to 
explore the subtle aspect of the quark mass dependence of the hyperfine splitting with a closer view. In the following section, we will see that including this effect reduces the $\chi^2/d.o.f.$ from 4.6 mentioned above to $\sim$1.0.

\subsection{Hyperfine splitting}\label{sec:hfs_func}

In the constituent quark potential model, the vector meson and the pseudoscalar mason are depicted
as $1^3S_1$ and $1^1S_0$ states, respectively, and their mass difference (the
hyperfine splitting) $\Delta_{\rm HFS}$ comes from the spin-spin contact interaction of the
valence quark and antiquark. A preliminary study on the behavior of $\Delta_{\rm HFS}$ with respect
to the quark mass $m_q$ on the lattice has been performed in Ref.~\cite{Dong:2007da,Li:2010pw}, 
where one finds that $\Delta_{\rm HFS}\propto 1/\sqrt{m_q}$ describes the data surprisingly well for $m_q$ ranging from the charm quark mass region down to almost the chiral region. (See Fig.~4 and Fig.~5 in
Ref.~\cite{Li:2010pw}.) For heavy quarkonium, this behavior can be understood qualitatively as
follows. In the quark potential model, the perturbative spin-spin interaction gives
\begin{equation}  \label{HFS}
\Delta_{\rm HFS}= \frac{16\pi\alpha_s}{9}\frac{\langle\mathbf{s}_1\cdot\mathbf{s}_2\rangle_{S=1}-
\langle\mathbf{s}_1\cdot\mathbf{s}_2\rangle_{S=0}}{m_Q^2}|\Psi(0)|^2,
\end{equation}
where $m_Q$ is the mass of the heavy quark, $\mathbf{s}_{1,2}$ are the spin operators of the heavy
quark and antiquark, and $\Psi(0)$ is the vector meson wave function at the origin. In view of the fact that charmonium and bottomonium have almost the same $2S-1S$ and $1P-1S$ mass splittings (N.B. this
equal spacing rule extends to light mesons as well, albeit qualitatively)
it is argued~\cite{Liu:1977et} that the size of the heavy quarkonium should scale as
\begin{equation}    \label{size}
r_{Q\bar{Q}}\propto \frac{1}{\sqrt{m_Q}}
\end{equation}
in the framework of the nonrelativistic Schr\"{o}dinger equation. This prediction is checked  
against the leptonic decay widths and the fine and hyperfine splittings~\cite{Liu:1977et} of
charmonium and upsilon and it holds quite well. Since $\Psi(0)$ scales as $(r_{Q\bar{Q}})^{-3/2}$, 
one finds from Eqs.~(\ref{HFS}) and (\ref{size}) that
\begin{equation}
\Delta_{\rm HFS}\propto \frac{1}{\sqrt{m_Q}}.
\end{equation}

\begin{figure}[htbp]
\begin{center}
\includegraphics[scale=0.7]{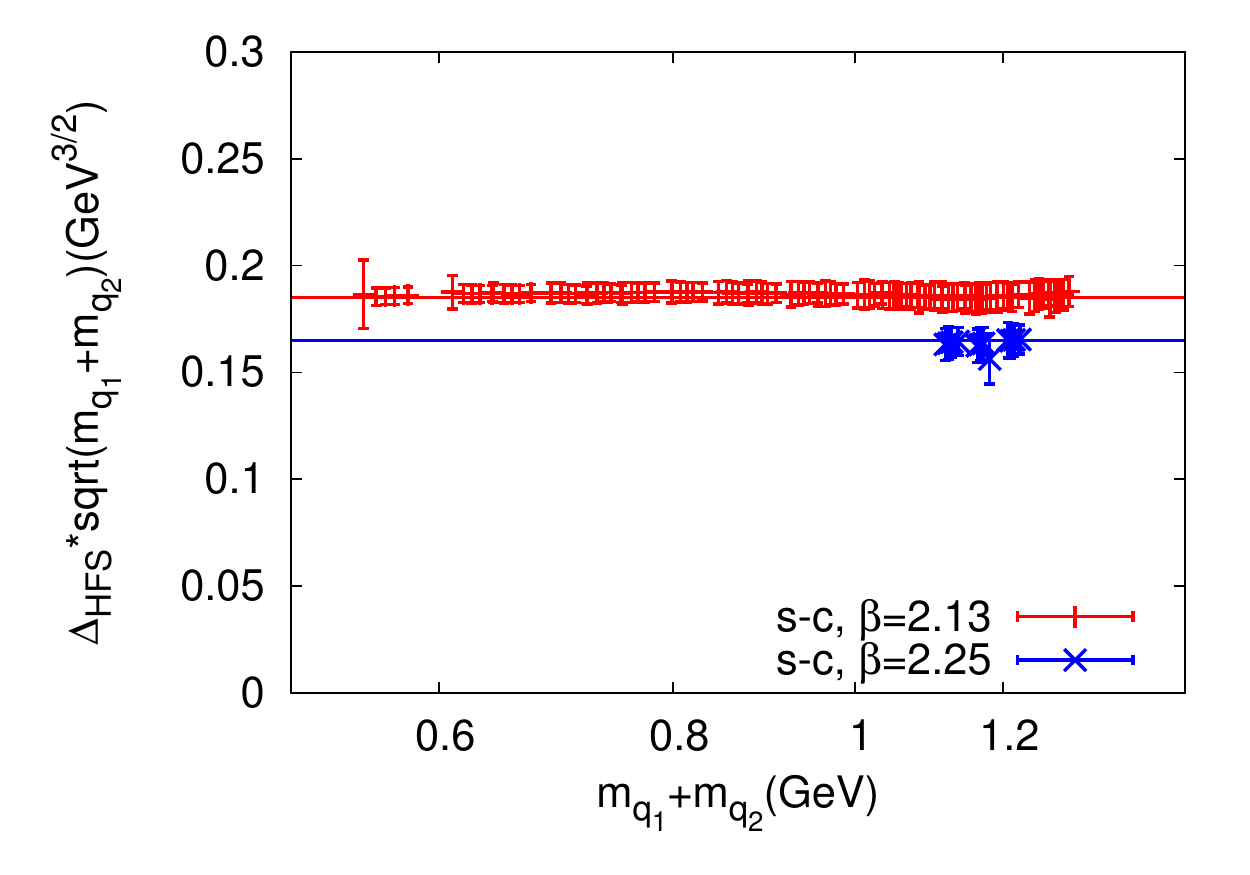}
\caption{The quark mass dependence of the hyperfine splittings $\Delta_{\rm HFS}=m_V-m_{\rm {PS}}$ times $\sqrt{m_{q_1}^R+m_{q_2}^R}$ for $\bar{c}{s}$ systems from the gauge ensemble with the lightest sea quark mass at $\beta=2.13/2.25$.
One can see that such a combination is consistent with a constant within one sigma for each ensemble, indicated by a solid line in the plot, obtained from an independent constant fit for each of the two ensembles.}\label{fig:splitting}
\end{center}
\end{figure}

\begin{figure}[htbp]
 \includegraphics[scale=0.7]{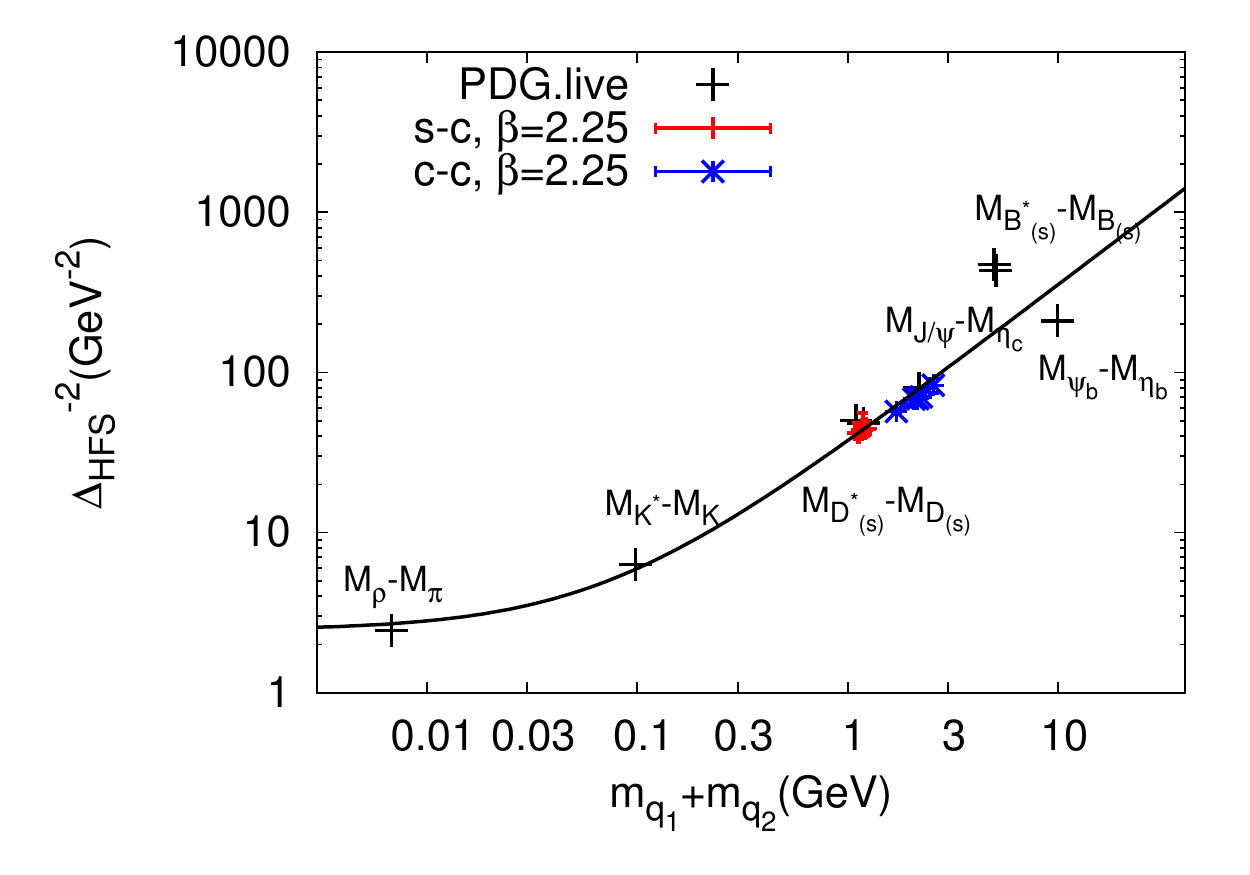}
 \caption{The $m_{q_1}+m_{q_2}$ dependence of the $\Delta_{HFS}^{-2}$, vs the physical quarks mass coming from PDG values~\cite{Agashe:2014kda} 
renormalized in $\overline{MS}$ scheme at 2 GeV. From left to right, the HFS for 
$M_{\rho}-M_{\pi}$,
$M_{K^*}-M_{K}$, $M_{D^{*}_{(s)}}-M_{D_{(s)}}$, $M_{J/\psi}-M_{\eta_c}$,
$M_{B^{*}_{(s)}}-M_{B_{(s)}}$ and $M_{\Psi_b}-M_{\eta_b}$ are plotted for 
comparison. The solid line in the plot is based on the correlated fit of the simulation data points with $\delta m$=0.068 GeV.}\label{fig:splitting_exp}
\end{figure}

Even though the above argument is for heavy quarkonium, it is interesting to see how far down
in quark mass it is applicable with a slight modification.
In the present study, we also check the quark mass dependence of $\Delta_{\rm HFS}$ for the
charm-strange systems. For clarity of illustration, 
the combined quantity $\Delta_{\rm HFS}\sqrt{m_{q_1}^R+m_{q_2}^R}$ from the gauge ensembles at $\beta=2.13/2.25$ with the lightest sea quark mass is plotted versus $m^R_{q_1}+m^R_{q_2}$ in Fig.~\ref{fig:splitting}, where one can
see that such a combination is consistent with a constant within one sigma in each ensemble, with $\chi^2/d.o.f.$=1.10 from the correlated fit with only the $1/\sqrt{m_{q_1}^R+m_{q_2}^R}$ term for all the data points in the six ensembles. The constants in the two ensembles in the Fig.~\ref{fig:splitting} are different which could be due to an $O(a^2)$ or $O(m_l)$ effect.
This suggests the following functional form
\begin{equation}\label{eq:hps-fit}
\Delta_{\rm HFS}=\frac{A_4+A_5 m_l^R}{\sqrt{m^R_{q_1}+m^R_{q_2}+\delta m}} (1+B_0 a^2).
\end{equation}
The parameter $\delta m$ is included since if $\delta m$ is zero, the hyperfine splitting will diverge in the chiral limit. With $\delta m=0$, the $\chi^2/d.o.f.$ is 0.87 which is better than the former fit without any $O(a^2)$ or $O(m_l)$ effect, and the $\chi^2/d.o.f.$ is almost the same when we set $\delta m\sim0.07$. Fig.~\ref{fig:splitting_exp} shows the $\Delta_{\rm HFS}^{-2}$ versus $m^R_{q_1}+m^R_{q_2}$ with the experimental data points from the review of the Particle Data Group in 2014~\cite{Agashe:2014kda}. The data of the charm-strange and charm-charm system in the ensemble with lightest sea quark mass at $\beta=2.25$ and the correlated fit of those data with $\delta m=0.068$ (the reason we choose this value will be discussed in Sec. \ref{sec:sys_err}) are also plotted on Fig.~\ref{fig:splitting_exp}. The fit we obtained could explain the splittings $M_{\rho}-M_{\pi}$,
$M_{K^*}-M_{K}$ within 10\% level, while the $B$ meson and bottomonium cases are beyond the scope of this form.

Finally, the global fit formula for the meson system is
\begin{eqnarray}\label{eq:para-cs-ori}
M(&\!\!\!m^R_c&\!\!\!,m^R_s,m^R_l,a)\nonumber\\ &=& \big[A_0 +A_1 m^R_c +A_2 m^R_s +A_3 m^R_l\nonumber\\
         &+& (A_4+A_5 m^R_l)\frac{1}{\sqrt{m^R_c+m^R_s+\delta m}}\big]\nonumber\\
   &\times& \big(1+ B_0 a^2 +B_1 (m^R_c a)^2 + B_2 (m^R_ca)^4\big)\nonumber\\
   &+& C_1 a^2
\end{eqnarray}
where $\delta m\approx$ 70 MeV is a constant parameter, the terms $A_5 m^R_l$ and $B_0 a^2$ are introduced for the light sea quark
mass and lattice spacing dependence of $\Delta_{\rm HFS}$.
 Note that $A_{2}$ is set to zero for the charm
quark-antiquark system, and $A_1$ is expected to be close to 1 (or 2) for the meson masses of
$\bar{c}s$($\bar{c}c$) system. 
We keep the $m^R_ca$ correction to the fourth order (${m_c^R}^4a^4\sim0.25$ for the physical charm quark mass at the ensembles at $\beta=2.13$, and just 0.0625 for the case at $\beta=2.25$), which turns out to
be enough (and necessary for the charmonium case) in the practical study.

In view of the observation from Fig.~\ref{fig:splitting} and Fig.~\ref{fig:splitting_exp} that the hyperfine splitting is primarily 
determined by the square root term, one expects that the parameters $A_0$, $A_1$, $A_2$ and $A_3$
of the corresponding pseudoscalar and vector meson masses to be the same within errors.

 It is true that one could do the polynomial expansion around the physical point, but then the fit would need to be iterated until the initial value for the center point of the polynomial expansion converges to the true physical point.  When we do this, we obtain consistent results compared to those of our preferred square-root fit, but the apparent simplicity of an interpolation is illusory.  With the square root term, we can skip the iteration, and, as a byproduct, have a possibly useful phenomenological form.

\vspace*{0.5cm}

\section{The Global fit and results}\label{sec:results}

Actually the meson masses measured from lattice QCD simulations are dimensionless values. Since we
will be determining the lattice spacing in a global fit, the fit formula in
Eq.~(\ref{eq:para-cs-ori}) cannot be used directly. Instead, we shall multiply the
lattice spacing $a$ to both sides of Eq.~(\ref{eq:para-cs-ori}) and modify the expression to

\begin{widetext}
\begin{eqnarray}
M a &=& \big[ A'_0 \frac{1}{C(a)} +A'_1 (m^R_c a) + A'_2 (m^R_s a) +A'_3 (m^R_l a)
                    + (A'_4+A'_{5} C(a)(m^R_l a))\frac{1}{C(a)^{3/2}} \frac{1}{\sqrt{(m^R_ca+m^R_sa+\delta m r_0/C(a))}} \big)\nonumber\\
   &&\times \big(1+ B'_0 \frac{1}{C(a)^2} +B'_1 (m^R_c a)^2 + B'_2 (m^R_c a)^4\big]  \nonumber\\
   && +C'_1 \frac{1}{C^3(a)}, \label{eq:para-cs-mod1}\\
\textrm{with }&&A'_0= A_0 r_0,\ A'_1=A_1,\ A'_2=A_2,\ A'_3=A_3,\ A'_4=A_4 r_0^{3/2},\ A'_5=A_{5} r_0^{1/2},\nonumber\\
                     &&B'_0=B_0 r_0^2,\ B'_1=B_1,\ B'_2=B_2,\ C'_1=C_1 r_0^3.\label{eq:para-cs-mod2}
\label{eq:para-cc-dim0}
\end{eqnarray}
\end{widetext} 
with $m_l^R$ fixed to the physical point (3.408(48) MeV) ~\cite{Bazavov:2009bb,Durr:2010vn,McNeile:2010ji,Laiho:2011np,Blum:2010ym,Kelly:2012uy}.
We have kept the $Ma$ and $m^R_q a$ combinations as they are, since $Ma$ is measured directly on the lattice and renormalized 
$m^R_q a$'s are used as the parameters of the sea quark and valence quark actions. For the explicit $a$'s which are not accompanied by a mass term, we have replaced them with $r_0/C(a)$ from Eq.~(\ref{eq:r0_over_a}).

Now, the global fit can be performed for all the relevant quantities using the
measured results from the six ensembles. It should to noted that the parameters to be fitted with
this expression are $A'_{0,1,2,3,4,5}$, $B'_{0,1,2}$, and $C'_1$ defined in Eq.~({\ref{eq:para-cs-mod2}}) for each physical quantity, and the universal parameter $\delta m$. 
For comparison, we have O(200) data points for  each physical quantity in the $\bar{c}s$ system (the total number of data points on all the six ensembles), and the corresponding number in the $\bar{c}c$ system is O(50).
Once we have fitted the coefficients $A'_{0,1,2,3,4,5}$ for 
the three dimensionless quantities $M_{D_s}a,  M_{D_s^*}a-M_{D_s}a$, and $M_{J/\psi}a$, we can examine their 
 dimensionful expressions,

\begin{eqnarray}
M_{D_s} &=& \frac{A^{'D_s}_0}{\bf r_0} +A^{'D_s}_1 {\bf m^R_c} +A^{'D_s}_2 {\bf m^R_s} +A^{'D_s}_3 m^R_l\nonumber\\
         &+& (\frac{A^{'D_s}_4}{\bf r_0^{3/2}}+\frac{A^{'D_s}_5}{\bf r_0^{1/2}} m^R_l)\frac{1}{\sqrt{{\bf m^R_c}+{\bf m^R_s}+\delta m}},\nonumber\\
\Delta_{\rm HFS, \bar{c}s} &=&  (\frac{A^{'\Delta}_4}{\bf r_0^{3/2}}+\frac{A^{'\Delta}_5}{\bf r_0^{1/2}} m^R_l)\frac{1}{\sqrt{{\bf m^R_c}+{\bf m^R_s}+\delta m}},\nonumber\\
M_{J/\psi} &=& \frac{A^{'J/\psi}_0}{\bf r_0} +A^{'J/\psi}_1 {\bf m^R_c} +A^{'J/\psi}_2 {\bf m^R_s}  +A^{'J/\psi}_3 m^R_l\nonumber\\
         &+& (\frac{A^{'J/\psi}_4}{\bf r_0^{3/2}}+\frac{A^{'J/\psi}_5}{\bf r_0^{1/2}} m^R_l)\frac{1}{\sqrt{{\bf m^R_c}+{\bf m^R_s}+\delta m}}.\nonumber\\
         \label{eq:para-cs-fit}
\end{eqnarray}
We see that they depend on the renormalized charm and strange quark masses $m^R_c$ and $m^R_s$,
and the scale parameter $r_0$ in the continuum limit. From the physical values of 
$M_{D_s}=1.9685$ GeV, $\Delta_{\rm HFS, \bar{c}s}\equiv M_{D_s^*}-M_{D_s}=0.1438$ 
GeV, and $M_{J/\psi}=3.0969$ GeV as inputs, we can determine $m^R_c, m^R_s$ and $r_0$. 

Note that the quark masses here are the ones renormalized under given scheme at given scale, specifically
$\overline{MS}(2 GeV)$ in our case. We ignore the tiny experimental uncertainties of these values. 

The use of the $J/\psi$ mass instead of the $\eta_c$ mass as one input is based on two considerations. 
First, the experimental $\eta_c$ mass is not as precisely determined as that for $J/\psi$. Secondly, the omission of the $\bar{c}c$ annihilation in the calculation of charmonium masses necessarily introduces systematic uncertainties. This kind of uncertainty is expected to be smaller for $J/\psi$ than for $\eta_c$~\cite{Levkova:2010ft}.

\begin{table*}[htpb]
\begin{center}
\caption{The fitting parameters (defined in Eq.~(\ref{eq:para-cs-ori})) for $M_{D_s}$, $M_{D_s^*}$ and $\Delta_{\rm HFS, \bar{c}s}\equiv M_{D_s^*}-M_{D_s}$. We list the ``default fit'' (keeping every parameter, listed as the first line of each channel), the ``optimal'' case (dropping the parameters which are consistent with zero, listed as the second line of each quantity), and also the parameters obtained in the global fit combining all the S-wave quantities (the third line of the $M_{D_s}$ and $\Delta_{\rm HFS, \bar{c}s}$ cases. $M_{D_s^*}$ doesn't have this line since we don't use it in the global fit). The $\chi^2$/d.o.f of first two cases are close, while the parameters from the optimal case have higher precision. }\label{table:two_fit}
\begin{ruledtabular}
\begin{tabular}{c|c|cccccccccc}
                      & $\chi^2/d.o.f.$&  $A_0$ & $A_1$ & $A_2$  & $A_3$ &  $A_4$ &  $A_5$ &$B_0$  & $B_1$ & $B_2$ & $C_1$\\
  \hline
$M_{D_s}$   &1.04                  & 1.343(140)& 0.881(34) & 0.791(25) & 0.28(3)  & -0.499(200) & 0.3(2)     & -0.14(8)    &0.07(10)       & 0.09(10)   &    0.11(13)  \\
                     &1.07                  & 1.200(54)  & 0.913(19) & 0.824(16) & 0.26(2)  & -0.338(36) & --              & --              &0.061(6)       &--              &   -0.18(2)  \\
                     &--                       & 1.297(23)& 0.891(8)    & 0.850(8)    & 0.30(2)  & -0.450(16) & --           & --              &0.057(3)       &  --            &    -0.18(1)  \\
  \hline
$M_{D_s^*}$&0.94                  & 1.17(8)      & 0.904(34) & 0.853(16) & 0.64(31) & -0.172(55)  & -0.07(33) & -0.10(8)   &0.088(34)     &-0.004(19)  &-0.01(13) \\
                     &0.95                  & 1.14(4)      & 0.913(16) & 0.840(14)   & 0.55(7)    &-0.137(18)   & --             & --             &0.070(6)      & --               &  -0.19(2)  \\
  \hline
$\Delta_{\rm HFS, \bar{c}s}$ 
                    &0.88                   &-0.025(18) & 0.02(3)     & 0.02(3)    & 0.5(5)      & 0.164(7)    & 0.23(15) &    0.36(16) & -0.36(22)    & 0.30(20)   & 0.07(4)\\
                    &0.91                   & --              &--               &--              &--              &0.158(7)     & 0.23(4)    &    0.35(12) &--                  &  --             &--  \\
                    &--                        & --              &--               &--              &--              &0.157(3)     & 0.29(6)    &    0.37(8)   &--                  &  --             &--  \\
\end{tabular}
\end{ruledtabular}
\end{center}
\end{table*}

We list in Table~\ref{table:two_fit} the fitting parameters (defined in Eq.~(\ref{eq:para-cs-ori})) for $M_{D_s}$, $M_{D_s^*}$ and $\Delta_{\rm HFS, \bar{c}s}\equiv M_{D_s^*}-M_{D_s}$. We give both the ``default fit'' (keeping every parameter, listed as the first line of each quantity) and the ``optimal'' case (dropping the parameters which are consistent with zero, listed as the second line of each channel). The $\chi^2$/d.o.f of two cases are close, while the parameters from the optimal case have higher precision. We note that the values of the coefficient $A_0$ of the constant term and the coefficients $A_1$, $A_2$, and $A_3$ of the terms with linear quark-mass dependence on $M_{D_s}$ obtained from the default fit are consistent with those obtained for the default fit of $M_{D_s^*}$  within errors. 
 
 So for the splitting $\Delta_{\rm HFS, \bar{c}s}$, these corresponding coefficients should be, and are, consistent with zero. To obtain results with higher precision we thus force these coefficients to be zero in our ``optimal'' fit for $\Delta_{\rm HFS, \bar{c}s}$.  In the first two rows of $\Delta_{\rm HFS, \bar{c}s}$ (the sixth and seventh rows from the top) in Table~\ref{table:two_fit}, we show that both the default fit with all the parameters (defined in Eq.~(\ref{eq:para-cs-ori}) as deduced from the combined parameters defined in Eq.~(\ref{eq:para-cs-mod1}) by using $r_0$ to be determined in the following section) and the optimal fit excluding the constant term and linear-quark-mass-dependence terms (thus keeping only the $1/\sqrt{m_q}$ term and its $O(a^2)$ corrections) are consistent, but the parameters we obtain from the latter one have higher precision. So compared with using $M_{D_s^*}$ as input, replacing it with the splitting $M_{D_s^*}-M_{D_s}$ gives more precise results for the predictions of the charm/strange quark mass and $r_0$. For the same reason, we discuss the splitting $\Delta_{\rm HFS, \bar{c}c}\equiv M_{J/\psi}-M_{\eta_c}$ instead of $M_{\eta_c}$ itself.

Note that we did the correlated fit for each quantity independently (turn on all the coefficients) and optimized the fit (turn off the negligible coefficients) to obtain the first two rows of each quantity in Table~\ref{table:two_fit}, then did the fully correlated global fit for all the $S$-wave quantities to obtain the third row of $M_{D_s}$ and $\Delta_{\rm HFS, \bar{c}s}$.  The case of $M_{D_s^*}$ doesn't have such a line since we don't use this quantity directly in the global fit. Due to the correlation between different quantities, in  Table~\ref{table:two_fit}, the parameters listed in the second line of  $M_{D_s}$ and $\Delta_{\rm HFS, \bar{c}s}$ are slightly different with that in the third line which are used for the final results and in the  following discussion.

\subsection{Systematic errors}\label{sec:sys_err}
 
In Tables~\ref{table:mass} and \ref{table:charm} we list both statistical and systematic errors.  For the statistical error we use the jackknife error of the global fit. Since we apply a global fit for data in all of six ensembles (two lattice spacings with three sea masses each), the errors from the $O(a^2)$ and $O(m_c^4a^4)$ corrections, and linear chiral extrapolation have been included in the statistical error.

For the systematic errors, we consider those concerning $r_0$, those of $Z_m(a)$, the global parameter $\delta m$, continuum/chiral extrapolation, the correlated fit cutoff, a possible electromagnetic effect, the effect from the missing charm sea, the one from the mixed action and the heavy quark data points in the ensembles at $\beta=2.13$.

\begin{enumerate}

  \item Since $r_0$ is the scale we want to determine in the global fit, we need to consider only two systematic errors: one from the statistical error of $C(a)=r_0(a)/a$, and the other from the non-zero $a^2$ dependence of $r_0(a)$.

   \begin{itemize}

     \item Our global fits use the central values for $C(a)$. The effect of the statistical errors of $C(a)$ for each value of $a$ used in the fit is incorporated into a systematic error as follows: For each lattice spacing, we repeat the global fit with the value of $C(a)$ changed by $1\sigma$ and calculate the resulting difference for each quantity of interest, namely $r_0$, $m_s$ and $m_c$, and then combine in quadrature the differences for each lattice spacing. This error will be marked with $\sigma(r_0/a)$.

     \item For simplicity, we constrain the fit parameter $r_0(a)$ at the physical point to be constant as a function of the lattice spacing in the global fits. But in principle there could be an $a$-dependence, with non-zero $c^a_1$ in Eq.~(\ref{eq:r0_new}).  In the work of RBC-UKQCD~\cite{Aoki:2010dy}, their fit gives $c^a_1=-0.25(14)$.  For such small $c^a_1$, the $\chi^2$ of the fit is almost unchanged: repeating the fit with $c^a_1=\pm0.25$ changes the $\chi^2/d.o.f.$ by 0.15\%.  For each quantity of interest, the change in its fit value is reported as a small systematic error in Table~\ref{table:mass}.  Had this been larger, it would have been incorporated into a statistical error instead, using $c^a_1$ as a fit parameter, but this was determined to be unnecessary {\it a posteriori}. This error will be marked with $\frac{\partial{r_0}}{\partial{a^2}}$. 

  \end{itemize}

  \item For the non-perturbative mass renormalization factor in the RI/MOM scheme, $Z_m(a)$, two kinds of systematic errors are involved.

  \begin{itemize}
  
    \item One is from the statistical errors of $Z_m(a)$. We follow the same procedure as the case of the 
 statistical errors of $C(a)$ to estimate the resulting effect on the quantities of interest: namely, for each lattice spacing we redo the fit with the value of $Z_m(a)$ changed by $1\sigma$, and then combine in quadrature the differences. For the strange and charm quark masses, this is the largest of the four systematic errors we tabulate.  A lot of this is due to the magnification of the statistical errors of $Z_m(a)$, which are independent at the two lattice spacings, upon extrapolation in lattice spacing.  This error will be marked with $\sigma$(MR/stat).

    \item On the other hand, the systematic errors of the perturbative matching from RI/MOM to the $
 \overline{MS}$ scheme, and the running of the mass renormalization factor to the scale of 2 GeV in $\overline{MS}$ scheme, are independent of lattice spacing and are totally the same for any simulation at any lattice spacing. Furthermore, since the physical quantities like meson mass or decay constant are independent of renormalization scheme or energy scale, this systematic error will not contribute to those quantities, only to the quark masses. For the quark masses, the systematic errors are independent of simulation and so are not magnified by linear extrapolation in lattice spacing.  These, then, are expected to be very small, which is what we see.  This error will be marked with $\sigma$(MR/sys).
    
    \end{itemize}
    
     \item Since the strange quark mass used in the domain-wall configurations ($\sim$120 MeV for the $\beta=2.13$ ensembles and $\sim$110 MeV for the $\beta=2.25$ ensembles) are not equal to the physical strange quark mass, a systematic error is induced.
     
     In Ref. \cite{Aoki:2010dy}, the reweighting of the strange quark mass is used to correct the values 
obtained from the original samples. 
In view of the fact that the strange sea quark mass has different values in the
two sets of the ensembles with different lattice spacing, it provides another way to estimate the systematic error from the mismatch of the strange sea quark mass.
     
     We can add the strange sea quark mass dependence terms into the functional form in Eq.~(\ref{eq:para-cs-ori}) with coefficients $A_6$ and $A_7$,
\begin{eqnarray}\label{eq:para-cs-reweight}
M(&\!\!\!m^R_c&\!\!\!,m^R_s,m^R_l,a) \nonumber\\
&=& \big[A_0 +A_1 m^R_c +A_2 m^R_s +A_3 m^R_l+{\bf A_6 m^R_{s,sea}}\nonumber\\
          &+& (A_4+A_5 m^R_l+{\bf A_7 m^R_{s,sea}})\frac{1}{\sqrt{m^R_c+m^R_s+\delta m}}\big]\nonumber\\
    &\times& \big(1+ B_0 a^2 +B_1 (m^R_c a)^2 + B_2 (m^R_ca)^4\big)\nonumber\\
   &+& C_1 a^2.
\end{eqnarray}
Since the form of the dependence of the lattice spacing and the strange sea quark mass are different, it is possible to distinguish them in the global fit. After we obtain the coefficients, the condition 
$m^R_s=m^R_{s,sea}$ is applied
to predict the Sommer scale $r_0$, the quark masses, and the other quantities. The $\chi^2$ of the fit with the strange sea quark mass extrapolation is almost the same as the one in the default case 
without such an extrapolation (1.06 vs 1.07), and the values of each quantity of interest in two ways of 
fit are consistent within error. 
We use the resulting difference of each quality of interest as the estimate of this systematic error. This error will be marked with $\sigma$(SSQMD). 

     \item In section \ref{sec:hfs_func}, we induce a parameter $\delta m\sim O(70)$ MeV since it provides better understanding of the light vector-pseudoscalar meson mass differences. In the correlated fit including all the S-wave related quantities, the value minimizes the $\chi^2$ is $\delta m=68$ MeV. To estimate the systematic error by this global parameter, we repeat the fit with $\delta m=68\pm 14$ MeV (20\% uncertainty) which changes the $\chi^2/d.o.f.$ by 1\% and the changes for the fit value of each quantity of interest is reported as a systematic error in Table.~\ref{table:mass}. This error will be marked with $\sigma(\delta m)$.

\begin{figure}[htbp]
 \includegraphics[scale=0.7]{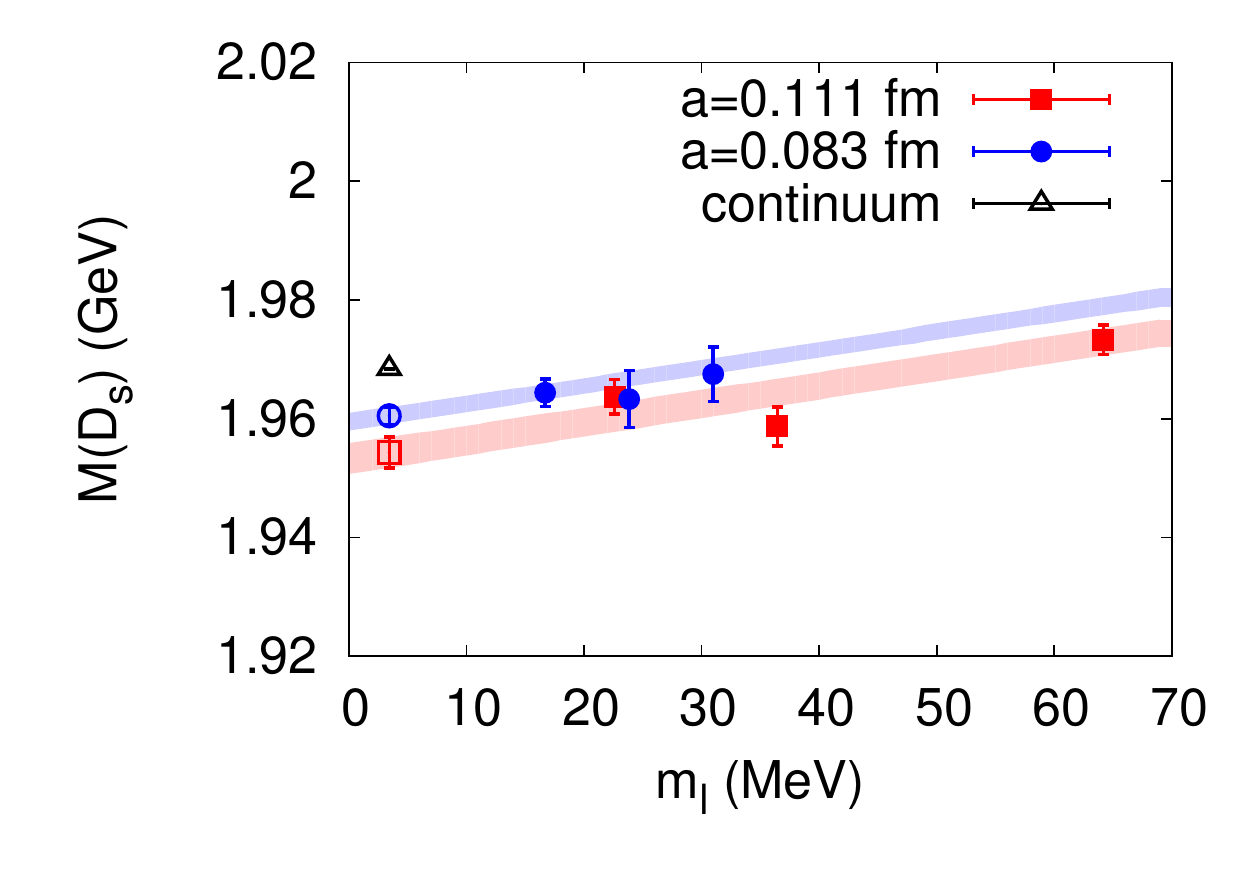}
 \includegraphics[scale=0.7]{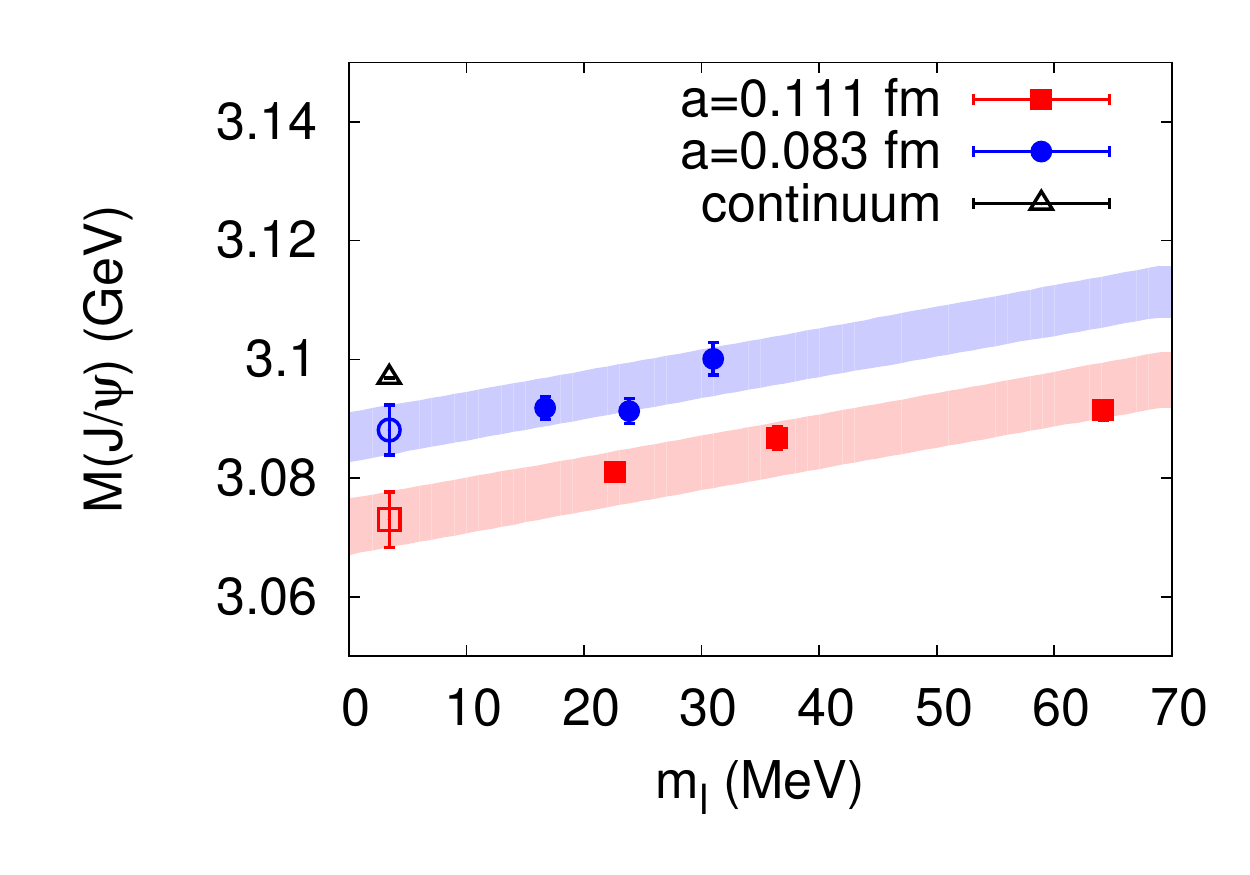}
 \caption{The interpolated values of the $M_{D_s}$ and $M_{J/\psi}$ from the data points of the two charm quark mass values which bracket the physical one, for each ensemble with different $\beta$, versus the renormalized  sea quark mass. The lattice spacing dependence of all the three quantities is manifest given the precision of the data. At the same time, the sea quark dependence of them is not negligible and trends similarly for each $\beta$.}\label{fig:input_depends}
\end{figure}  
     
     \item After we obtain the quark mass $m^R_{c,s}$ and Sommer scale $r_0$, we can do the interpolation on the data points of the two neighboring charm/strange quark masses, and plot in Fig.~\ref{fig:input_depends} the interpolated values for the $M(D_s)$ and $M(J/\psi)$ versus the renormalized  sea quark mass, for each ensemble with different $\beta$.  The error bands of the correlated fit and linear extrapolation of the sea quark mass in the different lattice spacings, and the continuum limit of them (the experimental inputs) are also plotted in the same figure. Most of the interpolated values are consistent with the fit, and the few exceptional ones reflect the statistical scatter from such a simple interpolation, compared to a global fit over a large quark mass region. 
     
     Fig.~\ref{fig:input_depends} shows the sea quark mass dependence of the interpolated values and that of the global fit, for the quantities we used as the inputs such as $M(D_s)$ and $M(J/\psi)$. It is obvious that the slope of the sea quark mass dependence is $\sim$0.5 from Fig.~\ref{fig:input_depends}  and not negligible, given the precision of the data points. Since the low lying meson mass in the charmonium system doesn't involved any valence light quark, there is no chiral perturbation theory available here to provide a reliable functional form of the sea quark mass dependence.  To estimate the systematic error from the chiral extrapolation, we added the $m_l^2$ term (following the twisted mass work \cite{Carrasco:2014cwa}) into the functional form of the charmonium mass quantities and repeated the fit, and took the changes for the fit value of each quantity of interest as a systematic error.
    
    On the other hand, a $m_K^3$ term involving valence strange quark and light sea quark could appear in the functional form of the $D_s$ or $D_s^*$ masses  due to the chiral perturbative theory,
like the $m_{\pi}^3$ term in the $K$ or $K^*$ masses. Since the valence strange quark mass is much heavier than the light sea quark mass, the effect of this term will not deviate far from a linear dependence of $m_l$. So we added the $m_K^3$ term into the functional form and took the change as a systematic error.
     
     For the decay constant of $f_{D_s}$, Ref. \cite{Sharpe:1995qp} shows a partially quenched form for its chiral behavior, 
\begin{eqnarray}
     f_{D_s}&=&c_0\big{[}1+ b_1m^2_{sl} \textrm{log}\frac{m^2_{sl}}{\Lambda_{QCD}^2} +  b_2 (m^2_{ss}-m^2_{ll}) \textrm{log}\frac{m^2_{ss}}{\Lambda_{QCD}^2}\nonumber\\
    && +c_1 m_s +c_2 m_l + ...\big{]}   
\end{eqnarray}
in which $m_{ss}$ is the mass of the valence pseudoscalar $\bar{s}s$, $m_{ll}$ is the pion mass in sea, and $m_{sl}$ is the valence-sea mixed kaon mass. We use the difference between this form and the trivial linear form as an estimate of the systematic error from the chiral extrapolation. 
     
     All the estimates of the error due to the chiral extrapolation form will be marked with $\sigma$(chiral).
     
     \item     For the possible $m_u-m_d$ effects, if the sea quark mass dependence of a quantity from the $u/d$ degenerated ensembles is 
\begin{eqnarray}
       A(m_l)=A_0+A_3 m_l,
\end{eqnarray}
     then we can rewrite it into 
\begin{eqnarray}
     A(m_l)=A_0+A^u_3 m_u+(A_3-A^u_3) m_d
\end{eqnarray}
     with a reasonable assumption $A^u_3\in (0, A_3)$ for the $u/d$ non-degenerate case. Then the upper bound of the non-degenerate effect happens at the boundary of the range of $A^u_3$, in extrapolating $m_l$ to $m_u$ (2.079(94) MeV) or $m_d$ (4.73(12) MeV), not the average of them 3.408(48) MeV\cite{Bazavov:2009bb,Durr:2010vn,McNeile:2010ji,Laiho:2011np,Blum:2010ym,Kelly:2012uy}. So the above estimate of the systematic error of the chiral extrapolation also includes the possible $m_u-m_d$ effects.  For all the quantities of interest, this effects are not larger than the standard estimate 0.2\% which comes from $(m_d-m_u)/m_p$.
     
     As seen in Fig.~\ref{fig:input_depends}, the slopes are close to each other,  so we can expect that the effect will not be large. This error will be marked with $\sigma$($u-d$). 
          
     \item Fig.~\ref{fig:input_depends} also shows that the lattice spacing dependence ($O(a^2)$) based on the functional form with $O(a^2)$ correction is obvious. In addition, $O(m_c^4a^4)$ dependence is not negligible in the $M(J/\psi)$ case. With the ensembles at only two values of lattice spacing, we can't justify the systematic error of such a lattice spacing dependence before we have the ensembles at $\beta>$2.25. But, if we change the functional form of the lattice spacing dependence in the chiral limit into
\begin{eqnarray}
M'(a)&=&A (1+B_1 m_c^2a^2+...)+ C_0 a^2 +O(a^4)\nonumber\\
&\sim&A(1+B_1 m_c^2a^2+...)/(1-\frac{C_0 a^2}{A})+O(a^4)\nonumber\\
&\sim&\bar{A}(1+B_1 m_c^2a^2+...)+\bar{C}_0 a^2+\frac{\bar{C}_0^2}{\bar{A}}a^4+O(m_c^2a^4))\nonumber\\
\end{eqnarray}
and solve $\bar{A}$ and $\bar{C}_0$ with the continuum limit at two lattice spacings, we find that the S-wave quantities are just changed by about 1 MeV and the changes of the P-wave quantities are a few MeVs. For each of the quantities of interest, we combine all the changes of the input quantities and the change of that quantity itself in quadrature, and treat it as a possible estimate of the systematic error of the lattice spacing dependence. This error will be marked with $\sigma(a)$. 
      
      \item Due to the precision of the data, the correlated fit require a cutoff for the small eigenvalue of the correlation matrix of the data points. The cutoff of the global fit is set to be $10^{-11}$, and we changed the cutoff into $10^{-11\pm1}$, repeated the fit, and took the changes for the fit value of each quantity of interest as the systematic error. This effect is very small in the cutoff region $10^{-12}$--$10^{-10}$ and the change of the $\chi^2/d.o.f$ is just 0.2\%. This error will be marked with $\sigma$(cut). 
      
      \item  As in reference \cite{Bazavov:2014wgs}, we can estimate the electromagnetic effect by modifying the mass of $D_s$ by 1 MeV. In our case, we used $M(D_s)$ and $M(D_s^*)-M(D_s)$ as the input, so we modified this two quantities by 1 MeV independently, and combined the changes in quadrature to estimate this systematic error. This error will be marked with $\sigma$(EM). 

     \item Our simulation is based on the 2+1 flavor domain-wall sea configuration which doesn't have any charm quark in the sea. Ref.~\cite{Levkova:2010ft} shows that without the disconnected charm diagram of the correlation function, the hyperfine splitting will decrease by a few MeV. But this affects only the mass of $\eta_c$ which is not the input quantity. So we think this effect will be negligible. 
     
%We repeat the global fit with the value of $M(J/\psi)$ changed by 1 MeV, and take the changes to estimate this systematic error.  This error will be marked with $\sigma(m_c^{sea})$. 

    \item The mixed action will inevitably introduce partial quenching. However, the low-energy constant $\Delta_{mix}$ for the overlap on RBC-UKQCD DWF 
configurations, as calculated by the combined DWF and overlap proapgators, is
very small \cite{Lujan:2012wg}. As a result, the mass of the mixed pion involving the light valence and sea which corresponds to $\sim 300$ MeV pion is only shifted by $\sim 10$ MeV on
the ensembles we used. 

The present work does not involve valence light quarks; we only calculate charmonium
and $\bar{c}s$ mesons. The only relevant terms from the $\chi$PT to the $\bar{c}s$ mesons are the non-analytic terms $m_K^3$ in the $D_s$, $D_s^*$ masses and the log$(m_K/\Lambda)$ term in $f_{D_s}$. We studied the effects of these terms and found that contributions from these terms in chiral extrapolation are up to one sigma of the statistical error which are included in the systematic errors as $\sigma$(chiral) in Table \ref{table:mass} and \ref{table:charm}. Since the mixed kaon mass is to be shifted by ~ 5 MeV only, this gives a deviation
of $\sim$3\% in $m_K^3$ and 0.15\% in log $(m_K/\Lambda)$. Such small changes on top
of very small contributions from the $\chi$PT terms other than the linear dependence 
of $m_l$ are not possible to discern and thus they are neglected in the present work. 

 The mixed action effect only appears in cases involving a valence-sea mixed meson such as exploring a scattering state with a two-valence-quark interpolation field, which is not relevant to this work, except for the decay constant of $D_s$ \cite{Sharpe:1995qp}. 

     \item As shown in Fig.~\ref{fig:eff_mass_ps}, the masses extracted for $am_c$ above $0.7$ could be suspicious, even when using the wall-source propagator. So a possible estimate of this systematic error is to remove the data points above $0.7$ from the analysis and determine how the result are affected. It turns out that this doesn't change the charm/strange quark mass and Sommer scale much, but affects other quantities by up to one sigma of the statistical error. 
     
     Another issue related to the $am_c$ error is that we forced the coefficient $B_0=0$ for all the quantities except the hyperfine splittings which we observed this effect clearly in Fig.~\ref{fig:splitting}. So we can turn on this coefficient in the global fit, repeat it and take the difference as the estimate of this error. Contrary tp the previous estimate, the change here affects the charm quark mass and the Sommer scale up to one sigma of the statistical error, and also slightly affects the other quantities.
     
     The error combined by the above two estimates in quadrature will be marked with $\sigma$(heavy). 

\end{enumerate}

\subsection{The charm and strange quark masses and Sommer scale
parameter $r_0$}\label{subsec:mass}
Our results for $m_{c}^{\overline{MS}}\textrm{(2 GeV)}$, $m_{s}^{\overline{MS}}\textrm{(2GeV)}$ and $r_0$ are listed in Table~\ref{table:mass}. The $\chi^2/d.o.f.$ of the fully correlated fit including $M(D_s)$, $\Delta_{HFS,\bar{c}s}$, $M_{J/\psi}$, $\Delta_{HFS,\bar{c}c}$ and $f_{D_s}$ is 1.05.

\begin{table}[htbp]
\begin{center}
\caption{The quark mass and Sommer scale $r_0$ after chiral and linear $O(a^2)$ extrapolation. The 
statistical error $\sigma$(stat), and twelve systematic errors from $r_0$ $\sigma$($r_0$/a) and $\sigma$($\frac{\partial{r_0}}{\partial{a^2}}$), the mass renormalization (MR) $\sigma$(MR/stat) and $\sigma$(MR/sys), the strange sea quark mass dependence $\sigma$(SSQMD), 
the parameter $\delta m$ $\sigma$($\delta m$), the chiral and continuum extrapolation $\sigma$(chiral)  and $\sigma(a)$, the possible $m_u-m_d$ effect $\sigma$($u-d$), the cut off of the correlated fit $\sigma$(cut), the electromagnetic effect $\sigma$(EM) and the heavy quark artifact $\sigma$(heavy) in the ensembles at $\beta=2.13$ are listed below the central values.}\label{table:mass}
\begin{ruledtabular}
\begin{tabular}{c|cccc}
                &  $\chi^2$/d.o.f & $r_0$(fm) & $m_{c}$(GeV)  & $m_{s}$(GeV)\\
 PDG~\cite{Agashe:2014kda}
                                  &                & --   & 1.09(3)       & 0.095(5)      \\
 \hline
this work                  & 1.05     & 0.465    &  1.118      & 0.101       \\
 $\sigma$(stat)       &              & 0.004    &  0.006      & 0.003      \\
 $\sigma$($r_0$/a)&             & 0.002    &  0.001       & 0.000       \\
 $\sigma$($\frac{\partial{r_0}}{\partial{a^2}}$)
                                 &              & 0.005    &  0.007        & 0.004       \\
 $\sigma$(MR/stat)&            & 0.001      &  0.022       & 0.000        \\
$\sigma$(MR/sys)  &            &  --            &   0.003      & 0.000        \\
$\sigma$(SSQMD)  &          &  0.006    &   0.004 & 0.002        \\
$\sigma$($\delta m$)&         &  0.001    &   0.001      & 0.000        \\
$\sigma$(chiral)      &            &  0.004    &   0.006      & 0.003        \\
$\sigma$($u-d$)      &            &  0.001    &   0.001      & 0.000        \\
$\sigma(a)$             &            &  0.002    &   0.002      & 0.001        \\
$\sigma$(cut)          &            &  0.001    &   0.001      & 0.000        \\
$\sigma$(EM)          &            &  0.002    &   0.002      & 0.001        \\
$\sigma$(heavy)      &           &  0.005    &  0.007      & 0.001        \\
$\sigma$(all sys)     &            & 0.009    &  0.024         & 0.006        \\                                   
$\sigma$(all)           &             & 0.010    &  0.025         & 0.007        \\
\end{tabular}
\end{ruledtabular}
\end{center}
\end{table}

\begin{figure}[htbp]
  \includegraphics[scale=0.7]{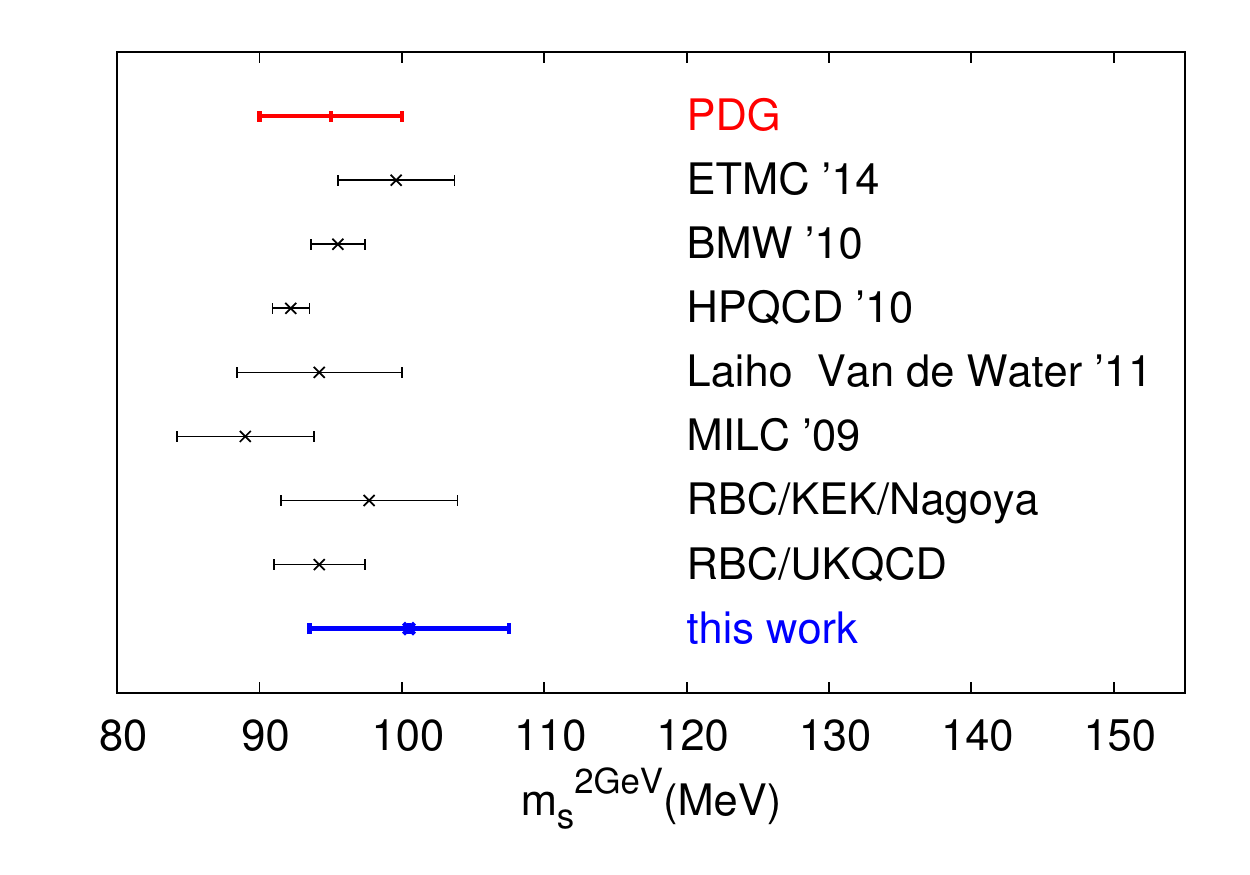}\\(a)\\
  \includegraphics[scale=0.7]{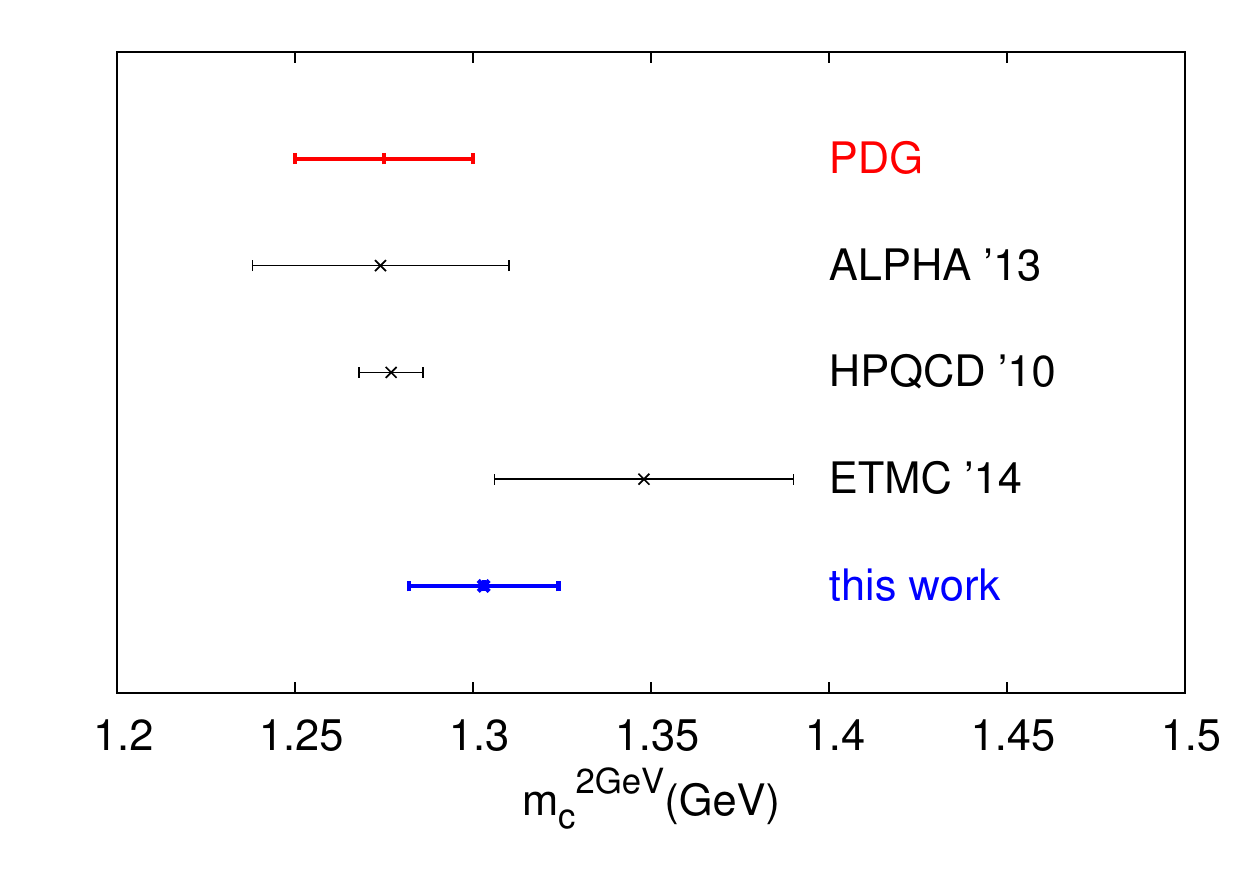}\\(b)
\caption{The prediction of $m_{s}^{\overline{MS}}$(2GeV)(upper panel) and $m_{c}^{\overline{MS}}(m_{c})$(lower panel) from this work, compared
to those of other works.}\label{fig:mass_compare}
\end{figure}

In Fig.~\ref{fig:mass_compare}(a), we plot our
results of $m_s^{\overline{MS}}$(2 GeV) to compare with the 2+1 flavor ones listed in lattice averages,  and another recent lattice calculation \cite{Carrasco:2014cwa}. The error bar in the plot and the following ones include both statistical and systematic errors. Since we determine the strange quark mass by the $\bar{c}s$ spectrum in which the strange quark mass only has a minor contribution, our result of $m_s$ is not quite precise, but it is consistent with the experimental data and the results of the other groups. Besides the statistical error, the systematic error from the $a^2$ dependence of the Sommer scale $r_0$, and that from the chiral extrapolation are as large as the statistical one, and contribute substantially to the total uncertainty. 

\begin{figure}[htbp]
  \includegraphics[scale=0.7]{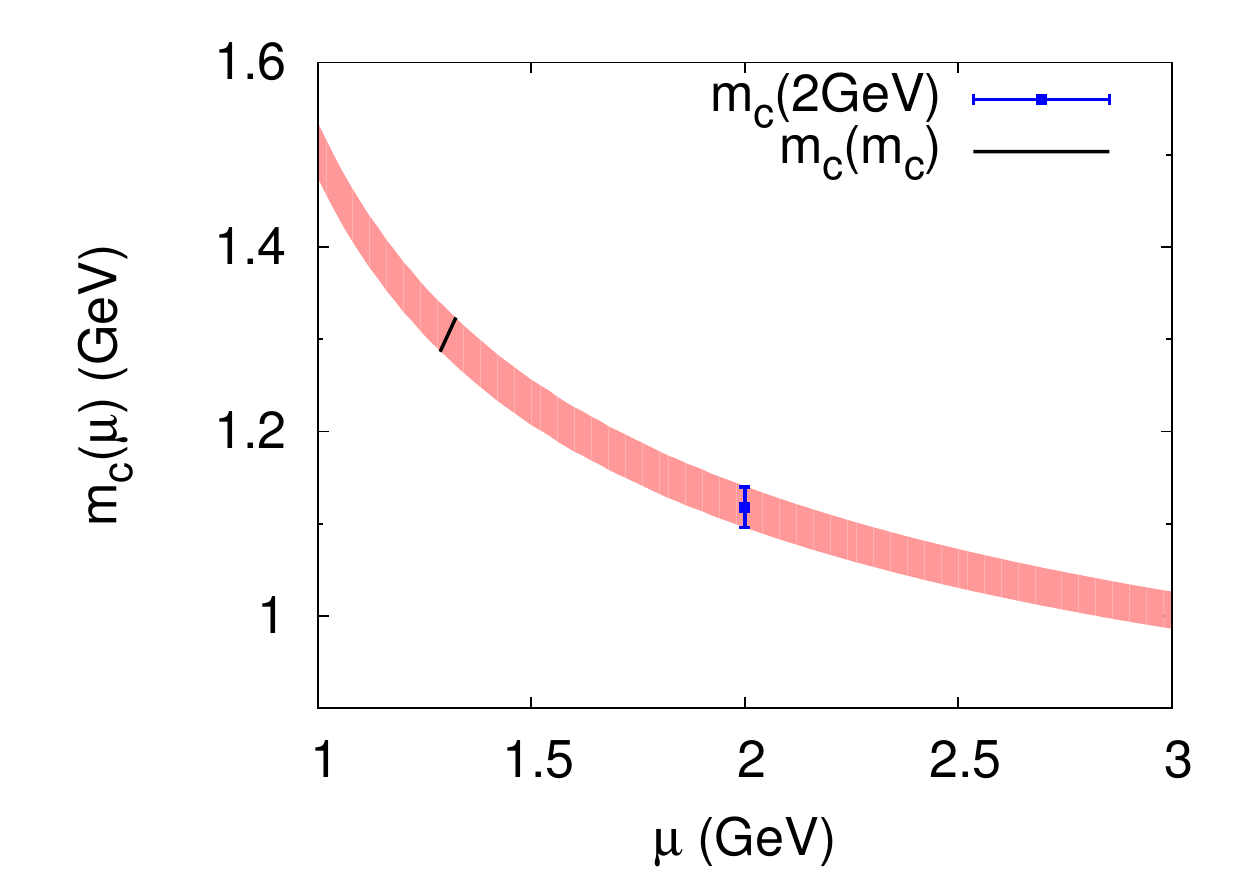}
\caption{The running charm quark mass $m_c(\mu)$ versus the scale $\mu$. Since the mass $m_c(m_c)$ is fully correlated to that scale, the uncertainty of $m_c(m_c)$ by the running from a given scale will be suppressed by approximately $\sqrt{2}$.}\label{fig:running}
\end{figure}

Our prediction of the value of $m_{c}^{\overline{MS}}$(2 GeV) is 1.118(6)(24) GeV. To obtain $m_{c}^{\overline{MS}}(m_c)$, we applied the quark mass running in reference \cite{Chetyrkin:1999pq,Xing:2007fb}. Note that the uncertainty of $m_{c}^{\overline{MS}}(m_c)$, indicated by the black band in Fig.~\ref{fig:running}, is not just the rescaling of the error at 2 GeV with the running factor from 2 GeV to the one of $m_{c}^{\overline{MS}}(m_c)$. It means that the error bar of $m_{c}^{\overline{MS}}(m_c)$ will be suppressed by $\sim\sqrt{2}$ compared to the estimate from naively rescaling. We repeat the running for the upper/lower band of $m_{c}^{\overline{MS}}$(2 GeV), obtain the on-shell scales of them, and average the changes comparing to the one of the central value of  $m_{c}^{\overline{MS}}$(2 GeV)  in quadrature as the estimate of the error of $m_{c}^{\overline{MS}}(m_c)$. The value of $m_{c}^{\overline{MS}}(m_c)$, 1.304(5)(20) GeV, is plotted in Fig.~\ref{fig:mass_compare}(b) to compare with those of ALPHA~\cite{Heitger:2013oaa}, HPQCD~\cite{McNeile:2010ji} and ETMC~\cite{Carrasco:2014cwa}.  
Considering the fact that HPQCD used $\cal O$(500) configurations per ensemble and that only $\cal O$(50 - 100) per 
ensemble are used in this work, the difference in precision between the results of HPQCD and this work
reflects the different statistics to a certain extent.

Note that the results of the strange/charm quark mass are based on the ensembles at only two lattice spacings and the systematics of the finite lattice spacing effect are not fully under control. We still need ensembles at least one more lattice spacing to access the full O($a^4$) errors.

\subsection{Charmonium spectrum}\label{subsec:charm}

Having determined the charm quark mass, we can predict the charmonium spectrum with the $J/\psi$ mass used as input.

There is a long story regarding the mass of $\eta_c$ in experiment and lattice calculation. In experiment, BELL~\cite{Abe:2007jna} and BES~\cite{Bai:2003et} obtained a value smaller than 2980 MeV about 10 years ago, while the BaBar result is around 2983 MeV. At present, all of their results~\cite{Vinokurova:2011dy,delAmoSanchez:2011bt,Ablikim:2012ur} are consistent with each other in the range 2982--2986, and the PDG average of the $\eta_c$ mass is 2983.7(7)~\cite{Agashe:2014kda}. 

\begin{figure}[htbp]
  \includegraphics[scale=0.7]{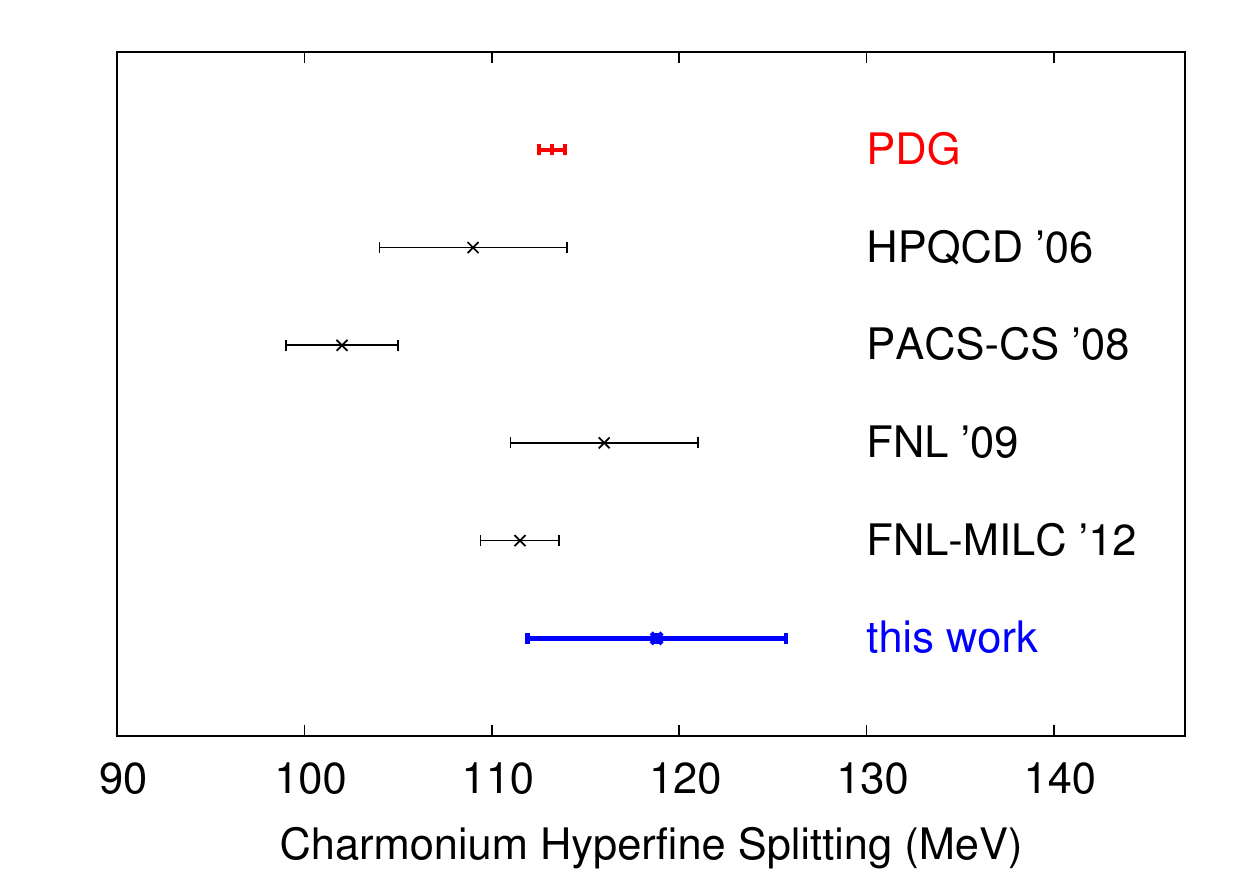}
\caption{The prediction of the hyperfine splitting of charmonium in this work, compared
to these of other works and experiment. Note the lattice results have not included the $\bar{c}c$ annihilation diagram which is expected to lower HFS by 1--5 MeV \cite{Levkova:2010ft,Feldmann:1998vh,Cheng:2008ss}.}\label{fig:sp_compare}
\end{figure}

In quenched lattice calculations, the hyperfine 
splitting result, namely the mass difference between $J/\psi$ and $\eta_c$, is much smaller than 
the physical value, only around 50--90 MeV, such as in Ref.~\cite{Allton:1992zy,Okamoto:2001jb,Choe:2003wx,Tamhankar:2005zi}. Such a difference is understood to be due to the effects of the shift of the coupling constant in a quenched simulation~\cite{Detar:2007ni}. A recent lattice result \cite{DeTar:2012xk} shows that the dynamical simulation could actually get a value close to 
experiment. At the same time, Ref.~\cite{Levkova:2010ft,Feldmann:1998vh,Cheng:2008ss} shows that without the disconnected charm diagram of the correlation function, the hyperfine splitting will increase by a few MeV.  So the correct 
lattice prediction of the hyperfine splitting should be slightly larger than the physical value for a dynamical 
simulation without the disconnected charm diagram. Our prediction of the value of the hyperfine splitting of charmonium, 119(2)(7) MeV, is plotted in Fig.~\ref{fig:sp_compare} to compare with experiment and other lattice results based on 2+1 flavor configurations. 

\begin{figure}[htbp]
 \includegraphics[scale=0.7]{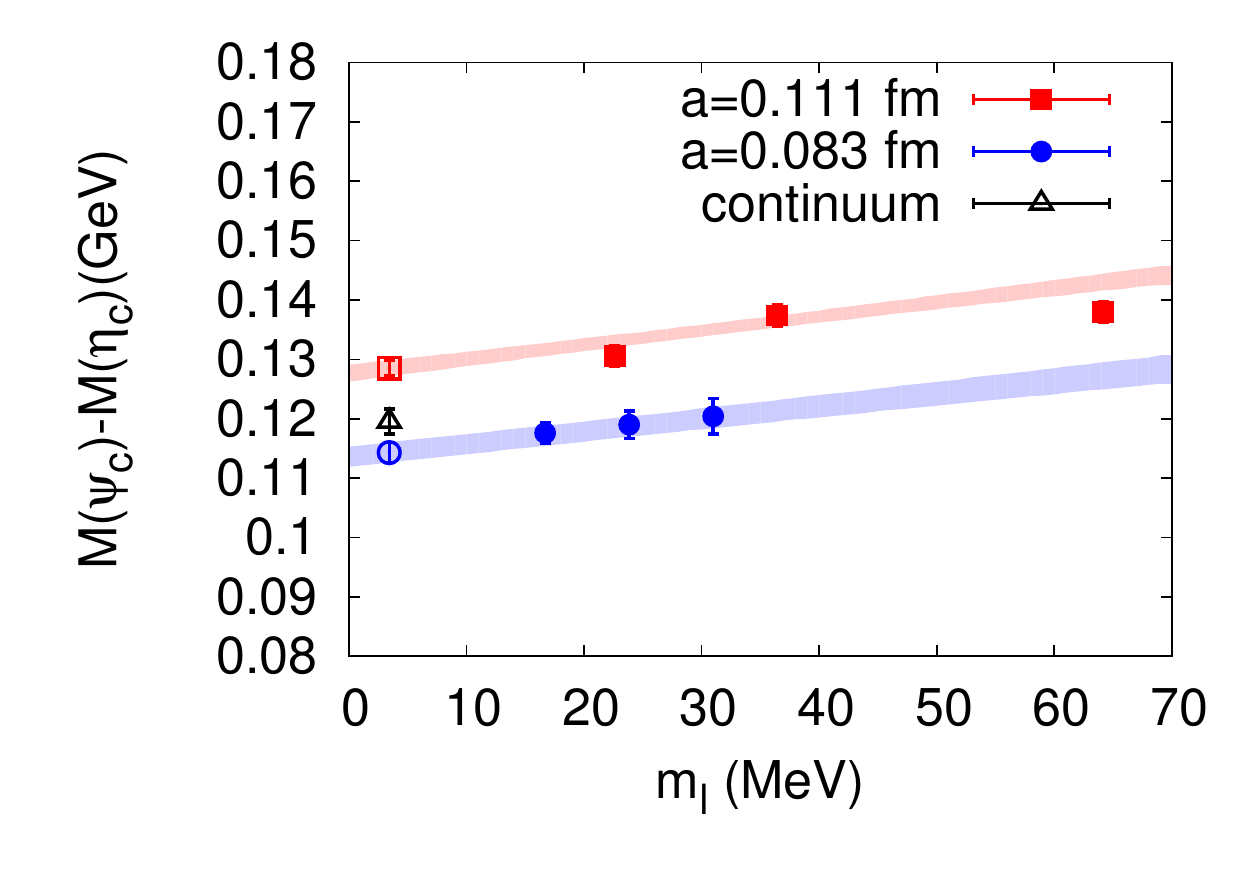}
 \caption{The interpolated values of $\Delta_{\rm HFS, \bar{c}c}$ on the data points of the two charm quark masses which bracket the physical one, for each ensemble with different $\beta$, versus the renormalized sea quark mass. Note that the $O(m^4a^4)$ effect is large so that then the continuum limit based on our functional form is between the data at the two finite lattice spacings.}\label{fig:cc_sp_depends}
\end{figure}

Fig.~\ref{fig:cc_sp_depends} shows the interpolated values of $\Delta_{\rm HFS, \bar{c}c}$, based on the data points of the neighboring two charm quark masses which bracket the physical one, for each ensemble with different $\beta$, versus the renormalized  sea quark mass. Note that the $O(m^4a^4)$ effect is large so that the continuum limit based on our functional form is between the chiral extrapolation at the two finite lattice spacings. So the present result would have an additional systematic error due to the functional form of the continuum extrapolation, and then would be changed somewhat if we had ensembles at $\beta>$ 2.25.

As mentioned in the beginning of Sec.~\ref{sec:results}, we fit the hyperfine splitting instead of the mass of $\eta_c$, and list it and its statistical and systematic uncertainty in Table~\ref{table:charm}. 

\begin{table}[htbp]
\begin{center}
 \caption{Charmonium spectrum results and $f_{D_s}$ after chiral and linear $O(a^2)$ extrapolation, in unit of GeV.}\label{table:charm}
\begin{ruledtabular}
\begin{tabular}{cccccc}
                  &$\Delta_{\rm HFS, \bar{c}c}$  & $M_{\chi_{c0}}$   & $M{\chi_{c1}}$   & $M_{h_{c}}$ & $f_{D_s}$\\
\hline                  
 PDG~\cite{Agashe:2014kda}
                                    & 0.1132(7)               & 3.4148(3)           & 3.5107(1)            & 3.5254(2)      & 0.258(6)\\
the work                     &  0.1188                  & 3.439                 & 3.524    &    3.518 & 0.2536\\
 \hline
 $\sigma$(stat)          &  0.0021                  & 0.037                 & 0.043    &    0.011  & 0.0022\\
 $\sigma$($r_0$/a)   &  0.0002                  & 0.001                 & 0.003    &    0.004  & 0.0001\\
 $\sigma$($\frac{\partial{r_0}}{\partial{a^2}}$)
                                      &  0.0018                 & 0.008                 & 0.034   &    0.056  & 0.0016\\
 $\sigma$(MR)           &  0.0008                  & 0.008                 & 0.009   &    0.039  & 0.0007 \\
 $\sigma$(SSQMD)   &  0.0027                  & 0.008                 & 0.009   &    0.005  & 0.0021 \\
  $\sigma$($\delta m$)& 0.0007                  & 0.001                 & 0.002   &    0.004  & 0.0005 \\
  $\sigma$(chiral)       &  0.0041                  & 0.001                 & 0.018   &    0.012  & 0.0006 \\
  $\sigma$($u-d$)       &  0.0000                  & 0.001                 & 0.000   &    0.008  & 0.0002 \\
 $\sigma(a)$               &  0.0008                 & 0.003                 & 0.004   &    0.003  & 0.0006 \\
  $\sigma$(cut)            &  0.0005                  & 0.000                 & 0.001   &    0.000  & 0.0007 \\ 
  $\sigma$(EM)           &  0.0008                  & 0.001                 & 0.003   &    0.004  & 0.0006 \\
$\sigma$(heavy)         &  0.0036                  & 0.018                 & 0.036   &    0.008  & 0.0025 \\
 $\sigma$(all sys)         &  0.0068                  & 0.023                & 0.052    &    0.070   & 0.0036\\                                       
$\sigma$(all)                &  0.0069                 & 0.044                & 0.066    &    0.071   & 0.0043\\
\end{tabular}
\end{ruledtabular}
\end{center}
\end{table}

Since  $P$-wave charmonium states are very noisy compared to the $S$-wave states, including them into the global fit will make the result quite unstable. So we don't include them in the global fit. Rather, we just do the correlated fit for the data points with different mass parameters, and use the quark mass and Sommer scale $r_0$ with their correlations as the inputs. Table~\ref{table:charm} also shows results for the mass of the $P$-wave charmonium states which are in good agreement with experiment.

Our prediction of $f_{D_s}$ shown in Table~\ref{table:charm} based on the global fit of the $S$-wave quantities will be discussed in the next section.

\subsection{Decay constant of $D_s$}\label{subsec:fds}

For a pseudoscalar (PS) meson, its decay constant $f_{PS}$ is defined through the hadronic matrix
element
\begin{eqnarray}
i\langle 0|\bar{s}\gamma_\mu\gamma_5 c|PS\rangle &=& f_{PS}p_{\mu},\label{eq:fds_1}
\end{eqnarray}
with $p_{\mu}$ the momentum of the PS meson.

Using the Ward identity of the partially conserved axial current (PCAC) \cite{Dong:2001fm}, the decay constant $f_{PS}$
could be also obtained by
\begin{eqnarray}
(m_{q_1}+m_{q_2})\langle 0|\bar{s} \gamma_5 c|PS\rangle=M_{PS}^2 f_{PS},\label{eq:fds_2}
\end{eqnarray}
with $M_{PS}$ being the mass of the PS meson. 

In a lattice simulation with local operators, the renormalization of the vector/axial-vector current is not equal to unity. So the $f_{PS}$ obtained from Eq.~(\ref{eq:fds_1}) requires the axial-vector renormalization factor to get the physical result:
\begin{eqnarray}
Z_A\langle0|\bar{\psi_a}\gamma_4\gamma_5\psi_b|PS\rangle^{bare}&=&M_{P}f_{PS}.\label{eq:fds_r_1}
\end{eqnarray}
On the other hand, the pseudoscalar current and mass renormalization involved in Eq.~(\ref{eq:fds_2}) are canceled ($Z_{PS}Z_m\equiv1$). This makes the 
$f_{D_s}$ from Eq.~(\ref{eq:fds_2}) free of the renormalization.

In this work, we construct four kinds of correlation functions 
\begin{eqnarray}
G_{A_4A_4}(t)&=&\langle \sum_{\overrightarrow{x}}\bar{s}(x)\gamma_4\gamma_5 c(x) \bar{c}(0)\gamma_4\gamma_5 s(0)\rangle\nonumber\\
G_{PA_4}(t)&=&\langle \sum_{\overrightarrow{x}}\bar{s}(x)\gamma_5 c(x) \bar{c}(0)\gamma_4\gamma_5 s(0)\rangle\nonumber\\
G_{A_4P}(t)&=&\langle \sum_{\overrightarrow{x}}\bar{s}(x)\gamma_4\gamma_5 c(x) \bar{c}(0)\gamma_5 s(0)\rangle\nonumber\\
G_{PP}(t)&=&\langle \sum_{\overrightarrow{x}}\bar{s}(x)\gamma_5 c(x) \bar{c}(0)\gamma_5 s(0)\rangle
\end{eqnarray}
to improve the precision of $f_{D_s}$. Combining the results of these four correlation functions, we
can get two kinds of matrix elements $\langle 0|\bar{s}\gamma_\mu\gamma_5 c|D_s\rangle$ and $\langle 0|\bar{s}\gamma_5 c|D_s\rangle$ required in Eq.~(\ref{eq:fds_r_1}) and Eq.~(\ref{eq:fds_2}), and then obtain  $f_{D_s}$. The vector/axial-vector renormalization factor required in Eq.~(\ref{eq:fds_r_1}) is 1.111(6) for the $\beta=2.13$ lattice and 1.086(2) for the $\beta=2.25$ lattice, as given in Ref.~\cite{Liu:2013yxz}.

We found that the average of two estimate of $f_{PS}$ (254(2)(4) MeV) obtained from Eq.~(\ref{eq:fds_r_1}) and (\ref{eq:fds_2}) provides a prediction consistent with those of the $f_{PS}$ obtained from these two equations separately (253(2)(5) and 255(3)(4)), while the $\chi^2$/d.o.f of the averaged $f_{PS}$ is smaller (0.8 for the averaged case vs. 1.2 for the two separated cases). The final result is listed in Table~\ref{table:charm}.

\begin{figure}[htbp]
  \includegraphics[scale=0.7]{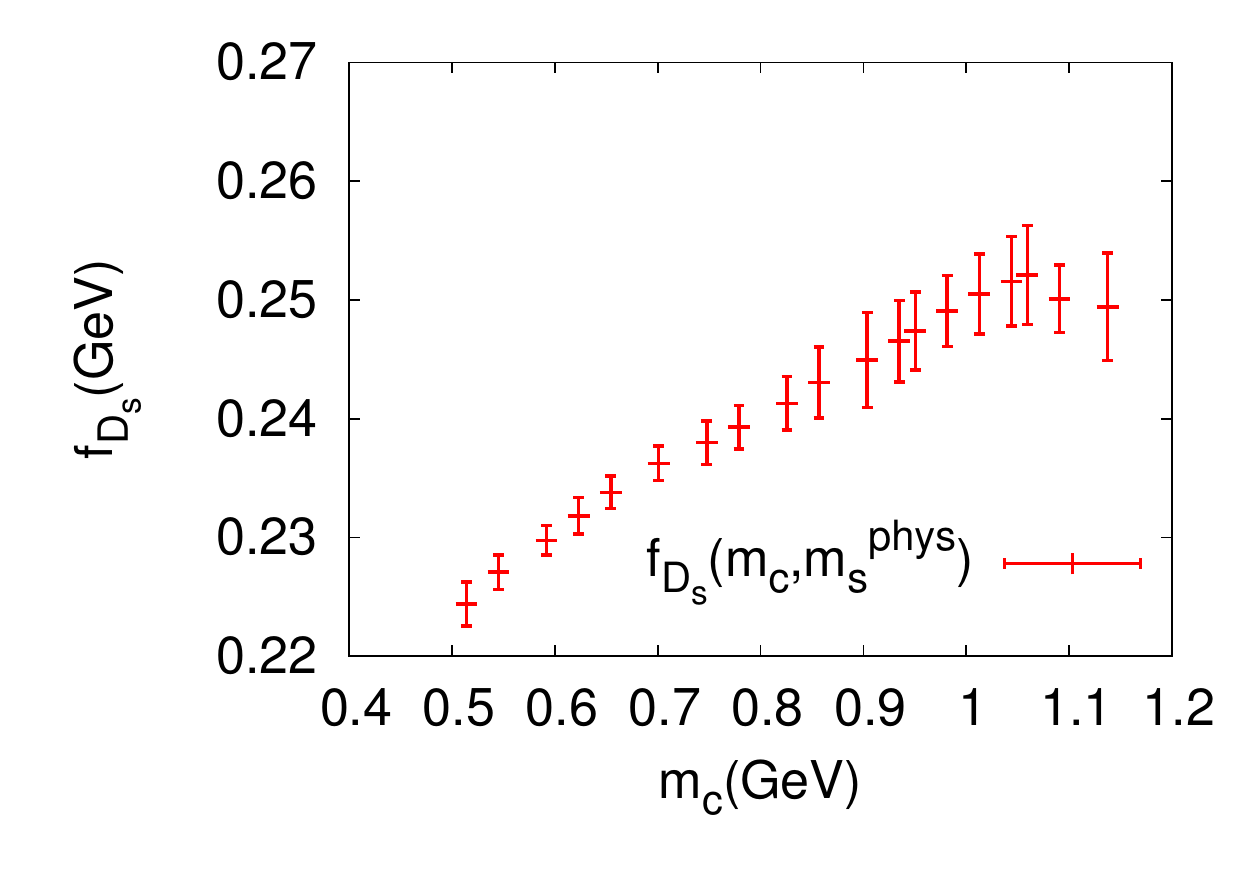}
    \includegraphics[scale=0.7]{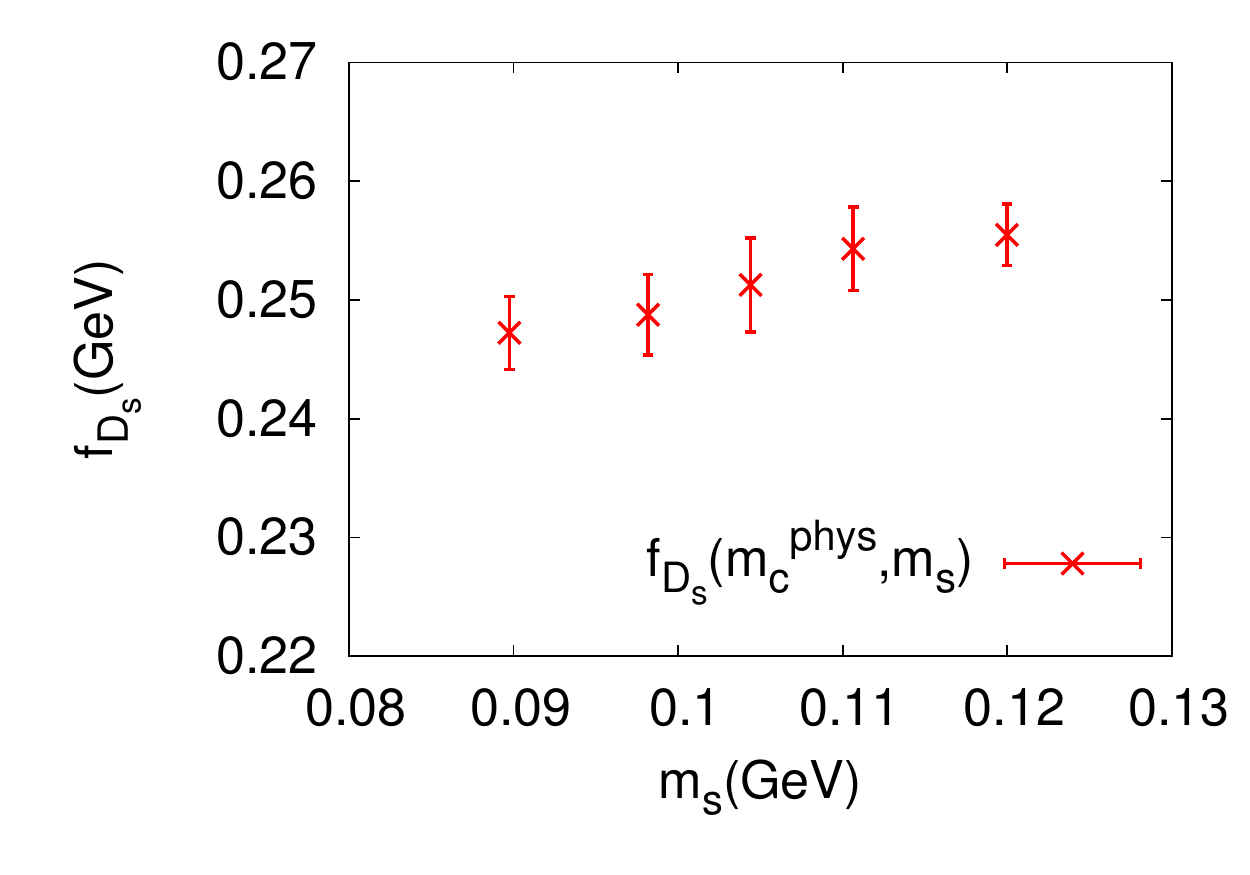}
\caption{The charm quark mass dependence (upper panel) and the strange quark mass dependence (lower panel) of $f_{D_s}$ with the other quark mass close to the physical point. This plot is based 
on the $\beta=2.13$ ensemble with lightest sea quark mass as an illustration.}\label{fig:f_ds_q_dep}
\end{figure}

It is interesting to show the charm/strange quark mass dependence on the $f_{D_s}$ in Fig.~\ref{fig:f_ds_q_dep}, in which the dependence of the charm quark mass is much stronger than that of the strange quark mass, when the other quark mass is fixed around the physical point. 

%The uncertainty of our prediction of $f_{D_s}$ mostly comes from the systematic error of the chiral extrapolation (about 1.0\%) and the statistical error (about 0.9\%); both the systematic error from the mismatch of strange sea quark mass and its physical value, and the heavy quark artifact in the ensembles at $\beta$=2.13 are around 0.8\%; that from the $a^2$ dependence of the Sommer scale $r_0$ is around 0.6\%. The effects from the other systematic errors are smaller than 0.3\%.

The uncertainty of our prediction of $f_{D_s}$ mostly comes from the statistical error (about 0.9\%); both the systematic error from the mismatch of strange sea quark mass and its physical value, and the heavy quark artifact in the ensembles at $\beta$=2.13 are around 0.8\%; that from the $a^2$ dependence of the Sommer scale $r_0$ is around 0.6\%. The effects from the other systematic errors are smaller than 0.3\%. Note that the systematic error due to the finite lattice spacing seems to be small based on the functional form of the continuum we used, but it is not fully under control since the simulation is based on the ensembles at just two lattice spacings.

\begin{figure}[htbp]
  \includegraphics[scale=0.7]{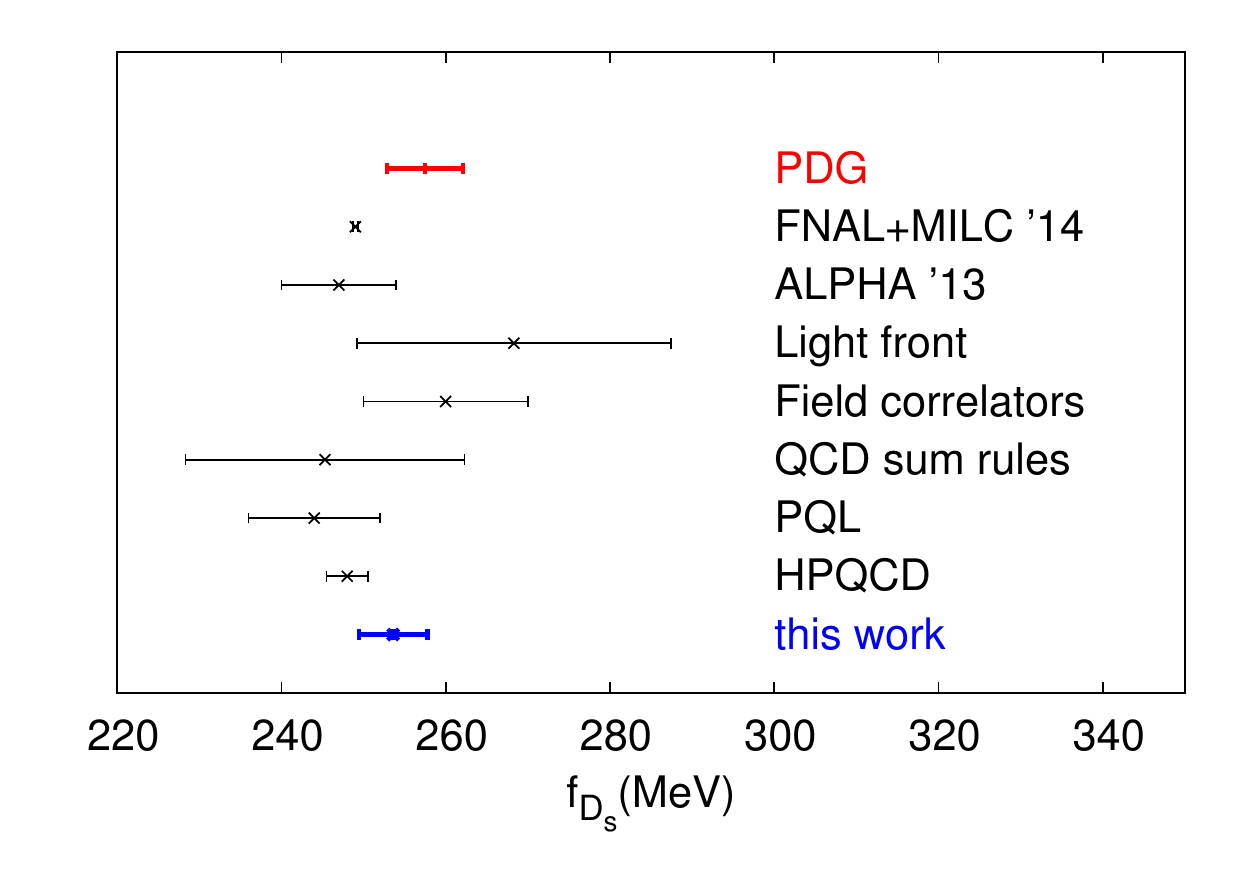}
\caption{The prediction of $f_{D_s}$ in this work, compared
to those of other works.}\label{fig:fds_compare}
\end{figure}

The comparison with other results of $f_{D_s}$ is illustrated in Fig.~\ref{fig:fds_compare}. 

\section{Conclusions}\label{sec:summary}

In this work, we used six ensembles of the 2+1 flavor gauge configurations with the Domain Wall sea quarks from RBC-UKQCD Collaboration, which include two lattice spacings 
each with three different light sea quark masses, to do the simulation for the spectrum of $\bar{c}s$  and $\bar{c}c$. With the global fit scheme, we can determine the charm/strange quark masses and Sommer scale $r_0$ using input from three physical quantities, $M_{D_s^*}$, $M_{D_s^*}-M_{D_s}$, and $M_{J/\psi}$.  Note that the results are based on the ensembles at only two lattice spacings and the systematics of the finite lattice spacing effect are not fully under control. 

Our prediction of the Sommer scale parameter 
\begin{eqnarray}
r_0=0.465(4)(9)\textrm{fm}
\end{eqnarray}
 is very close to the one obtained by HPQCD (0.4661(38)fm), and the one
determined by RBC-UKQCD (0.48(1)). With the $r_0$ obtained here, the lattice spacing of the 
$\beta$=2.13 and 2.25 ensembles
are 0.112(3) and 0.084(2) fm respectively
(or 1.75(4) and 2.33(5) GeV$^{-1}$ respectively).

The strange/charm quark masses we obtain are
\begin{eqnarray}
&m_{s}^{\overline{MS}}\textrm{(2 GeV)}&=0.101(3)(6) \textrm{GeV},\nonumber\\
&m_{c}^{\overline{MS}}(m_{c})&=1.304(5)(20) \textrm{GeV},\nonumber\\
&m_c^{\overline{MS}} ({\rm 2 GeV}) &= 1.118(6)(24) \textrm{GeV},\nonumber\\
\textrm{and,}\nonumber\\
&\frac{m_{c}^{\overline{MS}}}{m_{s}^{\overline{MS}}}\textrm{(2 GeV)}&=11.1(0.8)
\end{eqnarray}
Both the strange and charm masses are consistent with their PDG averages~\cite{Agashe:2014kda} 
which includes many calculations from the lattice simulation. 

For the charmonium hyperfine splitting, our result 
\begin{eqnarray}
\Delta_{\rm HFS, \bar{c}c} =119(2)(7) {\rm MeV}
\end{eqnarray}
is consistent the PDG average of 113.7(7) MeV~\cite{Agashe:2014kda}. Considering the 
possible effect of the disconnected diagram ($\sim$1--5 MeV)~\cite{Levkova:2010ft,Feldmann:1998vh,Cheng:2008ss}, our prediction could be smaller by one sigma, and thus even better in agreement. Besides the hyperfine splitting, we also checked the mass spectrum of the P-wave mesons, $M_{\chi_{c0}}$=3.439(44) GeV, $M_{\chi_{c1}}$=3.524(66) GeV, and $M_{h_{c}}$=3.518(71) 
GeV. The uncertainty of all of them are at the 2\% level and the values are in agreement with experimental results 3.4148(3) GeV, 3.5107(1) GeV and 3.5254(2) GeV, within one sigma.

\begin{figure}[htbp]
\includegraphics[scale=0.7]{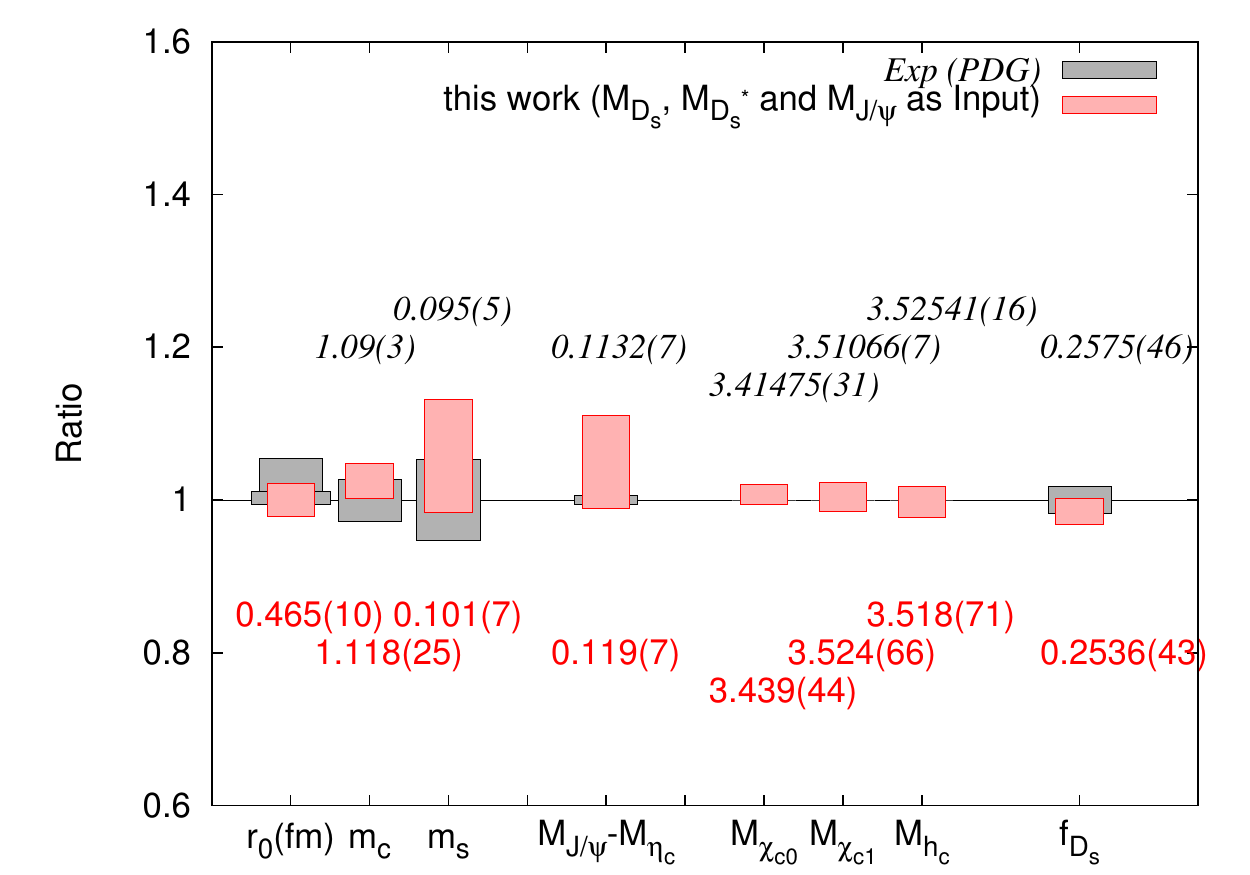}
\caption{We list the ratio of our simulation results to PDG averages. Note that the numbers of PDG
averages are in italic type, and all the numbers are in unit of GeV, except $r_0$. For $r_0$, we list
its value from HPQCD (0.4661(38) fm) and RBC-UKQCD (0.48(1) fm) for reference. Note that the values of the renormalized charm and strange quark masses are those at 2 GeV in $\overline{MS}$ scheme.}\label{fig:ratio}
\end{figure}

Another important prediction of this work is that of $f_{D_s}$. Our result
\begin{eqnarray}
 f_{D_s}=254(2)(4) {\rm MeV}
\end{eqnarray} 
is in agreement with experiment at 257.5(4.6) MeV, and other lattice simulations and phenomenology calculations.
The ratio of our results for various quantities to their corresponding PDG averages~\cite{Agashe:2014kda}
 are plotted in Fig.~\ref{fig:ratio},
to provide a direct comparison of their consistency.

The calculation in this work is based on configurations at two lattice spacings. We still need ensembles at least one more lattice spacing to access the full $O(a^4)$ errors, and lighter sea quark masses closer to those of the physical ones, to confirm their systematic effects. Besides that, reducing the systematic error from the strange sea quark being not at the physical point and including the disconnected charm diagram, could result in better estimates.

\section*{Acknowledgments}
We thank the RBC and UKQCD Collaborations for making their DWF configurations available to us. This work is supported in part by the National Science Foundation of China (NSFC) under Grants
No. 11335001, No. 11075167, No. 11105153 and also by the U.S. Department of Energy under Grant
No.~DE-FG05-84ER40154. Y.C. and Z.L. also acknowledge the support of NSFC under
No. 11261130311 (CRC 110 by DFG and NSFC). A.A. acknowledges the support of NSF CAREER through grant PHY-1151648. Z.L. is partially supported by the Youth Innovation Promotion Association of CAS. We thank the Center for Computational Sciences of the University of Kentucky for their support.

 \end{document}